\documentclass[12pt]{article}
\usepackage{epsfig}
\usepackage{amssymb}
\usepackage{cite}
\usepackage{rotating}

\newcommand\sss{\scriptscriptstyle}
\newcommand\sQ{{\sss Q}}

\newcommand\nlf{{n_{\rm\scriptscriptstyle L}}}
\newcommand\nf{{n_{\rm\scriptscriptstyle f}}}
\newcommand\tf{{T_{\rm\scriptscriptstyle F}}}
\newcommand\cf{{C_{\rm\scriptscriptstyle F}}}

\newcommand\aem{\alpha_{\rm em}}
\newcommand\as{\alpha_{\rm\scriptscriptstyle S}}

\def\asb{{}\ifmmode \bar{\alpha}_s \else $\bar{\alpha}_s$\fi}
\newcommand\MSB{$\overline{\rm MS}$}
\newcommand\mQ{m}
\newcommand\LambdaQCD{\Lambda_{\scriptscriptstyle \rm QCD}}

\newcommand\muF{\mu}
\newcommand\muz{\mu_0}

\def\beq{\begin{equation}}
\def\eeq{\end{equation}}

\def\beqn{\begin{eqnarray}}
\def\eeqn{\end{eqnarray}}
\def\lq{\left[}
\def\rq{\right]}
\def\rg{\right\}}
\def\lg{\left\{}
\def\({\left(}
\def\){\right)}
\def    \nn             {\nonumber}

\def\Dnp{D_{\rm NP}}
\def\pDnp{\tilde{D}_{\pi}}
\def\gDnp{\tilde{D}_{\gamma}}

\def\ord#1{{\cal O}\(#1\)}

\def\s{\sigma}

\def\Re{\mathop{\rm Re}}
\def\mthr{m_{\rm thr}}

\relax

\relax

\begin{document}
\begin{titlepage}
\nopagebreak
{\flushright{
        \begin{minipage}{4cm}
        Bicocca-FT-05-23 \\
        LPTHE-05-24
        \end{minipage}        }

}
\vspace{1cm}
\begin{center}
{\LARGE 
{ \bf \sc
A Study of Heavy Flavoured Meson\\
 Fragmentation Functions\\
in $e^+e^-$ annihilation

}}
\vskip .5cm
{\bf Matteo Cacciari}
\\
\vskip .1cm
{LPTHE, Universit\'e Pierre et Marie Curie (Paris 6), France}

\vskip .5cm
{\bf Paolo Nason}
\\
\vskip 0.1cm
{INFN, Sezione di Milano\\
Piazza della Scienza 3, 20126 Milan, Italy} \\

\vskip .5cm
{\bf Carlo Oleari}
\\
\vskip 0.1cm
{Universit\`a di Milano-Bicocca,\\
Piazza della Scienza 3, 20126 Milan, Italy} \\
\end{center}
\nopagebreak
\begin{abstract}
We compare QCD theoretical predictions for heavy flavoured mesons
fragmentation spectra in $e^+e^-$ annihilation with data from CLEO, BELLE
and LEP. We include several effects in our calculation:
next-to-leading order initial conditions, evolution and coefficient
functions. Soft-gluon effects are resummed at next-to-leading-log
accuracy. A matching condition for the crossing
of the bottom threshold in evolution is also implemented
at next-to-leading order accuracy.
Important initial-state electromagnetic radiation effects in the
CLEO and BELLE data are accounted for.
We find that, with reasonably simple choices of a non-perturbative
correction to the fixed-order initial condition for the evolution,
the data from CLEO and BELLE can be fitted with remarkable accuracy.
The fitted fragmentation function, when evolved to LEP energies,
does not however represent fairly the 
$D^*$ fragmentation spectrum measured by ALEPH.
Large non-perturbative corrections
to the coefficient functions of the meson spectrum are needed
in order to reconcile CLEO/BELLE and ALEPH results.\newline
Non-perturbative parameters extracted from the fits to $e^+e^-$ fragmentation
data for $D/D^*$ and $B$ mesons are tabulated. They can be employed 
in the theoretical predictions for the production of charmed and bottomed
mesons in 
hadron-hadron, photon-hadron and photon-photon collisions.
\end{abstract}
\vskip 1cm
\tableofcontents
\vfill
\end{titlepage}

\section{Introduction}

The study of the fragmentation functions of heavy flavoured hadrons
is of considerable interest in several aspects of QCD and collider
physics. On one hand, at transverse momenta much larger than
the mass of the heavy flavour, the heavy-flavour fragmentation functions
behave similarly to the light-hadron ones,
and obey an Altarelli-Parisi evolution equation. Unlike the case of
light hadrons, however, heavy-flavour fragmentation is very hard,
and thus probes a region of the evolution equation in the large
$x$ regime. Furthermore, it is dominated by the non-singlet component, at
least at 
moderate energies, so that its study is much simpler.
The initial condition for evolution has a well defined perturbative
expansion in powers of $\as(\mQ)$, $\mQ$ being the mass of the
heavy flavour, and a well defined expansion in terms of Sudakov
(i.e.\ large-$x$) logarithms. Further corrections of non-perturbative
origin, suppressed by powers of the ratio $\LambdaQCD/\mQ$,
can be parametrized to attempt to give a uniform description
of charm and bottom fragmentation functions.
These parameterizations
can then be employed to provide theoretical predictions for 
charmed and bottomed hadron production in hadron-hadron, photon-hadron
and photon-photon collisions at large transverse momenta.

Recently, new, high quality data on charmed meson production
have come from CLEO~\cite{Artuso:2004pj} and BELLE~\cite{Seuster:2005tr}.
One thus has the opportunity
to perform a more accurate fit to the non-perturbative
initial conditions, and furthermore one can test the evolution
of the fragmentation function from centre-of-mass energies 
of 10.6 to 91.2~GeV, using charm data from LEP experiments.
In the present work we will carry out this program.
We develop a procedure that overcomes various difficulties
in the large-$x$ region of the fragmentation function.
We are thus able to fit the measured fragmentation functions using
next-to-leading logarithmic (NLL) evolution, next-to-leading
order (NLO) initial conditions, NLO coefficient functions, NLL
Sudakov resummation 
(both for the initial conditions and for the coefficient functions) and
a phenomenological non-perturbative component. 
With a suitable choice of this non-perturbative component
we can obtain very good fits to CLEO and BELLE
data for $D^*$ and $D$ fragmentation, over the whole $x$ range. 
All the moments of the fragmentation functions
are therefore well reproduced. This represents an improvement 
over previous investigations
where, while obtaining good fits to some low moments
(a necessary and sufficient
condition for predicting heavy-meson production in hadronic collisions)
the fit in $x$-space was not completely satisfactory.
The same procedure is also applied to
$B$ meson spectra measured in $Z^0$ decays.
The relevant data from  the ALEPH~\cite{Heister:2001jg}
OPAL~\cite{Abbiendi:2002vt}, SLD~\cite{Abe:2002iq} and
DELPHI~\cite{Baker:2002,Ben-haim:2004kn} collaborations
are equally well described.

We evolve the $D^*$ fragmentation function
fitted to CLEO and BELLE data up
to LEP energies,
taking into account the opening of the bottom threshold~\cite{Cacciari:2005ry}.
We find a discrepancy between the QCD prediction and the ALEPH
data~\cite{Barate:1999bg}, that can be parametrized as a power correction of the
form $C(N-1)/E$, with $C$ of the order of few hundreds MeV,
or of the form $C(N-1)/E^2$, with $C\approx 5~{\rm GeV}^2$, where
$E=\sqrt{q^2}$ 
is the total centre-of-mass energy and $N$ is the moment in Mellin space.
Unfortunately, there is no way, at the moment, to discriminate
between the two possibilities. Theoretical arguments based upon
renormalons disfavour the presence of $1/E$ corrections in the
evolution of fragmentation functions. On the other hand, these
arguments require validation.

In Section~\ref{sec:thframe} we describe the
theoretical ingredients that enter our calculation, collecting and summarizing
previously  available results: the perturbative initial
condition, the evolution, the Sudakov effects and the bottom threshold. The
novel treatment of the large-$N$/large-$x$ region is also detailed here.
In Section~\ref{sec:isr} we describe our
treatment of electromagnetic initial-state radiation.
The implementation of the non-perturbative component of the fragmentation
function 
is discussed in Section~\ref{sec:nonpFF}.
In Section~\ref{sec:cleobelle} we perform fits to the
CLEO and BELLE data. In Section~\ref{sec:aleph} we compare
the evolved CLEO/BELLE $D^*$ fit to the ALEPH data,
and discuss in detail the related problems.
Fits for $B$ meson production are presented in Section~\ref{sec:bfit}.
In Section~\ref{sec:moment},
we perform fits to data under a different perspective, using Mellin moments 
rather than the $x$-space distributions, and employing a
simpler, one-parameter non-perturbative function, that can be related
to $\Lambda/m$ power corrections.
The implications of the new BELLE and CLEO data for heavy-flavour
hadroproduction 
are also discussed in this section.
Finally, in Section~\ref{sec:Conc}, we give our conclusions.

\section{Theoretical framework}
\label{sec:thframe}
We consider the inclusive production of a heavy quark
$Q$ of mass $m$
\beq 
\label{eq:process}
  e^+ e^- \,\to\, Z/\gamma\;(q) \,\to\, Q\,(p) + X\;,
\eeq
where $q$ and $p$ are the four-momenta of the intermediate boson and of the
final quark.
We define $x$ as the scaled energy
of the final heavy quark,
\beq
   x \equiv \frac{2\, p\cdot q}{q^2}\;.
\eeq
In our framework, we neglect corrections suppressed by powers of the
heavy-quark (meson)  mass, and so the above definition may be
replaced with the usual experimental definition of scaled momentum
(i.e.\ the heavy flavoured meson momentum over its maximum value).
The inclusive cross section for the production of the heavy quark Q can 
be written as a perturbative expansion in $\as$
\beq
\label{eq:sigma}
   \frac{d\sigma_{\sss Q}}{dx}(x,q^2,\mQ^2)
= \sum_{n=0}^{\infty} \asb^n(\mu^2)\, \sigma^{(n)}_{\sss Q}(x,q^2,\mQ^2,\mu^2)\;,
\eeq
where $E=\sqrt{q^2}$ is the total centre-of-mass energy, $\mu$ is the
renormalization scale, and 
\beq
 \asb(\mu^2) \equiv  \frac{\as(\mu^2)}{2\pi}\;.
\eeq
The cross section~(\ref{eq:sigma}), normalized to the total cross section, is
sometimes referred to as the heavy-quark fragmentation function in $e^+e^-$
annihilation.

In order not to spoil the convergence of Eq.~(\ref{eq:sigma}), the
coefficients $\sigma^{(n)}_{\sss Q}$ should be small enough to justify a
perturbative 
expansion in terms of $\as$. There are, however, 
at least two interesting regions of the parameter phase space
where such convergence is undermined:
\begin{enumerate}
\item 
If $q^2\gg \mQ^2$, large logarithms of the form $\log(q^2/\mQ^2)$ appear in
the 
differential cross section~(\ref{eq:sigma}) to all orders in the perturbative
expansion. These logarithms have collinear origin, and the mass $\mQ$ acts as
a regulator. 

\item In the region of the phase space of multiple soft-gluon emission, i.e.\
  $x\,\to\, 1$, the differential cross section contains enhanced terms
proportional to $\log^n(1-x)/(1-x)$. 
\end{enumerate}

In the following two sections, we collect the relevant formulae
for the resummation of these large
contributions at the next-to-leading log level.

\subsection{Collinear logarithms}\label{sec:Collinear_logarithms}
In the limit where power terms of the ratio $\mQ^2/q^2$ can be neglected, the
differential cross section satisfies the factorization theorem 
\begin{equation}
\label{eq:factorization}
   \frac{d\s_{\sss P,Q}}{dx}(x,q^2,\mQ^2) =
 \sum_i \int_x^1 \frac{dz}{z}\, 
 C_{{\sss P},i}\(z,q^2,\muF^2\) \,
  D_i\(\frac{x}{z},\muF^2,\mQ^2\) \;,
\end{equation}
where the subscript ${\rm P}$ stands for either T for transverse, L for longitudinal,
A for asymmetric or nothing for the total (i.e.\ L$+$T) cross
section\footnote{We  follow closely the notation of Ref.~\cite{Nason:1994xx}.}.
In the following, we shall always drop the polarization subscript, since we shall
always refer to total cross sections.

The $C_{i}$ coefficients are the \MSB-subtracted partonic cross
sections for producing the massless parton $i$, and $D_i$ are the
\MSB\ fragmentation functions for parton $i$ to evolve
into the heavy quark $Q$. The factorization scale
$\muF^2$  must be taken of order $q^2$ in
order to avoid the appearance of large logarithms of $q^2/\muF^2$ in the
partonic cross section.  The explicit expressions for the partonic
cross sections and for the fragmentation functions at NLO
can be found in Refs.~\cite{Mele:1990cw,Nason:1994xx}.

The \MSB\ fragmentation functions  ${D}_i$ obey the Altarelli-Parisi
evolution equations\footnote{Notice that the splitting functions
are transposed with respect to the structure function evolution equations.}
\beq
\label{eq:AP}
  \frac{d D_i}{d\log\mu^2} (x,\mu^2,\mQ^2) = \sum_j\int^1_x \frac{dz}{z}\,
   P_{ji}\(\frac{x}{z},\asb(\mu^2)\) \;D_j(z,\mu^2,\mQ^2)\;.
\eeq
The Altarelli-Parisi splitting functions $P_{ji}$ have the perturbative
expansion
\beq
\label{eq:APfunctions}
    P_{ji}\Bigl(x,\asb(\mu^2)\Bigr) = \asb(\mu^2) P^{(0)}_{ji}(x) 
    + \asb^2(\mu^2) P^{(1)}_{ji}(x) + {\cal O}(\asb^3)\;,
\eeq
where the $P_{ji}^{(0)}$ are\footnote{
The $+$
distribution is defined as
\begin{eqnarray*}
\label{eq:plus_distrib}
\int_0^1 dz\;h(z) \;[\, g(z) \,]_+  \equiv \int_0^1 dz \;[h(z)-h(1)]\;g(z)
\;.
\nonumber
\end{eqnarray*}
}~\cite{Altarelli:1977zs}
\begin{eqnarray}
P^{(0)}_{qq}(z)&=&C_{\sss F}\left[\frac{1+z^2}{(1-z)_+}
+\frac{3}{2}\delta(1-z)\right]\;,
\nonumber \\
P^{(0)}_{gg}(z)&=&2 C_{\sss A}\left[
\frac{z}{(1-z)_+}+\frac{1-z}{z}+z(1-z)+
\left(\frac{11}{12}-\frac{\nf T_{\sss F}}{3 C_{\sss A}}\right)
\delta(1-z)\right]\;,
\nonumber \\
P^{(0)}_{gq}(z)&=&C_{\sss F}\frac{1+(1-z)^2}{z}\;,
\nonumber \\
P^{(0)}_{qg}(z)&=& T_{\sss F}\left[z^2+(1-z)^2 \right]\;,
\end{eqnarray}
$\nf$ is the number of active flavours and
\begin{equation}
C_{\sss A}=3,\quad C_{\sss F}=\frac{4}{3}, \quad T_{\sss F}=\frac{1}{2}\;.
\end{equation}
The NLO splitting functions $P_{ji}^{(1)}$ (needed to achieve NLL accuracy) 
have been computed in
Refs.~\cite{Curci:1980uw,Furmanski:1980cm,Floratos:1981hs,Kalinowski:1980we,Kalinowski:1980ju}.
They are too lengthy to be replicated here.

The initial conditions for the \MSB\ fragmentation functions
were first obtained at the NLO level in Ref.~\cite{Mele:1990cw}.
They are given by
\beqn
\label{eq:Dini_Q}
{D}_\sQ(x,\muz^2,\mQ^2)&=&\delta(1-x)+\asb(\muz^2)\,
 d^{(1)}_\sQ(x,\muz^2,\mQ^2)+{\cal O}(\asb^2)\;,\\ 
\label{eq:Dini_g}
{D}_g(x,\muz^2,\mQ^2)&=&\asb(\muz^2)\, d^{(1)}_g(x,\muz^2,\mQ^2)+{\cal
  O}(\asb^2)\;,
\eeqn
(all the other components being of order $\asb^2$), where
\beqn
d^{(1)}_\sQ(x,\muz^2,\mQ^2)&=&\cf\left[
\frac{1+x^2}{1-x}\left(\log\frac{\muz^2}{\mQ^2}-2\log(1-x)-1\right)\right]_+
  \,,
\\
d^{(1)}_g(x,\muz^2,\mQ^2)&=& \tf \lq x^2+(1-x)^2 \rq\log\frac{\muz^2}{\mQ^2}\;.
\eeqn

In order to compute the NLL resummed fragmentation function, one takes the
initial conditions at a scale $\muz\simeq \mQ$,
evolves them up to $\mu\simeq E$ (these choices for $\muz$ and $\mu$
prevent the appearance  of large logarithms that would spoil
the NLL accuracy), and then
applies Eq.~(\ref{eq:factorization}), using the NLO expression for
the partonic cross sections $C_{i}$ given in Eqs.~(2.15)
of Ref.~\cite{Nason:1994xx}\footnote{In this work, we complement the Born
electroweak cross section with a threshold factor for the heavy quarks (and
antiquarks)
\begin{eqnarray*}
\sigma_{0,q}(q^2) \to \sigma_{0,q}(q^2) \left(1 +
\frac{2m_q^2}{q^2}\right)\sqrt{1 - \frac{4 m_q^2}{q^2}}\,,
\nonumber
\end{eqnarray*}
for $q=c,b$. Its numerical impact is, however, negligible at
the energies considered here.
}
\begin{eqnarray}
C_{q}(z,q^2,\mu^2) &=& \left[\delta(1-z)+\asb
  {a}^{(1)}_{q}\(z,\frac{\mu^2}{q^2}\) 
\right]\sigma_{0,q}(q^2) \;,\\
C_{g}(z,q^2,\mu^2) &=& \asb {a}^{(1)}_{g}\(z,\frac{\mu^2}{q^2}\)
  \sigma_{0,g}\;(q^2)\;, 
\end{eqnarray}
where, to make contact with the notations of
Refs.~\cite{Mele:1990cw,Nason:1994xx}, we have defined
\begin{equation}
a^{(1)}_{q/g}\(z,\frac{\mu^2}{q^2}\)
 \equiv \cf c_{q/g}\(z,\frac{\mu^2}{q^2}\)\;.
\end{equation}
The procedure outlined above guarantees that all leading 
and next-to-leading logarithmic terms of quasi-collinear origin
(terms of the form
$(\asb \log(q^2/\mQ^2))^n$ and $\asb (\asb \log(q^2/\mQ^2))^n$ respectively)
are correctly resummed in the final cross section.

When dealing with the type of convolution appearing in
Eqs.~(\ref{eq:factorization}) and~(\ref{eq:AP}), it is 
customary to introduce the Mellin transform
\beq
   f(N) \equiv \int_0^1 dx \, x^{N-1} f(x)\;.
\eeq
We adopt the convention that, when $N$ appears instead of $x$ as the
argument of a function, we are actually referring to the Mellin transform
of the function. The Mellin transform of the factorization 
formula~(\ref{eq:factorization}) is given by
\beq
\label{eq:sigmaN}
\sigma_{\sss Q}(N,q^2,\mQ^2) =\sum_i C_{i}(N,q^2,\muF^2) \;
{D}_i(N,\muF^2,\mQ^2)\;, 
\eeq
where
\beq
\sigma_{\sss Q}(N,q^2,\mQ^2) \equiv \int_0^1 dx \, x^{N-1}
 \frac{d\sigma}{dx}(x,q^2,\mQ^2)\;,
\eeq
and the Mellin transform of the Altarelli-Parisi
 evolution equation~(\ref{eq:AP}) at NLO is
\beq
\label{eq:APN}
 \frac{dD_i(N,\mu^2,\mQ^2)}{d\log\mu^2} = \sum_j 
 \asb(\mu^2) \left[ P_{ji}^{(0)}(N) + \asb(\mu^2) P_{ji}^{(1)}(N)\,
      \right] D_j(N,\mu^2,\mQ^2)\;.
\eeq
\subsection{Soft logarithms}\label{sec:Soft_logarithms}
Both ${a}_{q}^{(1)}$ and $d^{(1)}_\sQ$ contain terms 
associated to the emission of a soft (and collinear) gluon,
These terms give rise to a large-$N$ growth of the corresponding
Mellin transforms
\begin{equation}
\label{eq:a_Q^1}
{a}_q^{(1)}(N,q^2,\muF^2)  = 
\cf \lq  \ln^2N 
+ \(\frac{3}{2}+2\gamma_E - 2\ln\frac{q^2}{\muF^2}\) \ln N  + \alpha_q 
+ \ord{1/N} \rq \!,
\end{equation}
\begin{equation}
\label{eq:d_Q^1}
d_\sQ^{(1)}(N,\mu_0^2,\mQ^2) =
 \cf \!\left[ - 2\ln^2N + 
2\!\left( \ln \frac{m^2}{\mu_0^2} - 2\gamma_E + 1\! \right)\!\ln N +
 \delta_\sQ  + \ord{1/N} \right]\! , 
\end{equation}
where $\gamma_E = 0.5772\dots$ is the Euler constant and
\beqn
\alpha_q &=& \frac{5}{6}\pi^2-\frac{9}{2}+ \gamma_E^2+\frac{3}{2}\gamma_E +
      \(\frac{3}{2}- 2\gamma_E \) \log\frac{q^2}{\muF^2} \; ,
\\
\delta_\sQ &=& 2 - \frac{\pi^2}{3} + 2\gamma_E - 2\gamma_E^2 - 
\( \frac{3}{2} - 2 \gamma_E \) \ln \frac{m^2}{\mu_0^2} \; .
\eeqn
In Ref.~\cite{Cacciari:2001cw}, the all-order resummation of the large $\(\ln
N\)$ contributions has been performed to  next-to-leading log accuracy, that
is, all logarithms of the form $\as^n \ln^{n+1}N$ (leading logarithms) and
$\as^n \ln^{n}N$ (next-to-leading logarithms) have been correctly resummed.
In the following we summarize the results of Ref.~\cite{Cacciari:2001cw}.

The Sudakov resummation factor for the $e^+e^-$ coefficient function can be
written as
\begin{equation}
\label{eq:Delta_Q_S}
\Delta_q^S(N,q^2,\muF^2) = \exp\lq \ln N \, g^{(1)}(\lambda) +
g^{(2)}(\lambda)\rq \;, 
\end{equation}
where
\begin{equation}
\label{g1funms} 
g^{(1)}(\lambda) = 
\frac{A^{(1)}}{\pi b_0 \lambda} \;
\bigl[ \lambda + (1-\lambda) \ln (1-\lambda) \bigr] \;,
\end{equation}
\begin{eqnarray}
\label{g2funms}
g^{(2)}(\lambda) 
&=& \frac{A^{(1)}  b_1}{2 \pi b_0^3}
\left[ 2\lambda + 2 \ln (1-\lambda) +  \ln^2 (1-\lambda) \right]
\\
&+&
\frac{\left( B^{(1)} -2 A^{(1)}\gamma_E \right)}{2 \pi b_0} \ln (1-\lambda)
\nn
\\\nn
&-&\frac{1}{\pi b_0} \left[\lambda + \ln (1-\lambda) \right] 
\left( \frac{A^{(2)}}{\pi b_0} - A^{(1)} \ln \frac{q^2}{\mu^2} \right) 
- \frac{A^{(1)}}{\pi b_0} \;\lambda \;\ln \frac{q^2}{\muF^2} \;,
\eeqn
and where
\beq
\label{betass}
b_0 = \frac{11 C_A - 4 T_F \nf}{12\pi}\;,\;\;\;\;\; b_1 =
\frac{17 C_A^2 - 10 C_A T_F \nf -6 \cf T_F \nf}{24\pi^2}
\eeq
are the first two coefficients of the QCD $\beta$-function, and
\beqn
\label{a1a2}
A^{(1)}&=&\cf\,, \;\;\; A^{(2)}= \frac{1}{2} \,\cf\;
K = \frac{1}{2}\,\cf\left[ C_A \left(\frac{67}{18}- \frac{\pi^2}{6}\right)
- \frac{5}{9}\nf \right] ,
\\
\label{b1} 
B^{(1)}&=& -\frac{3}{2}\, \cf \,. 
\eeqn
The variable $\lambda$ is defined by
\beq
\label{eq:lambda}
\lambda \equiv b_0 \;\as(\mu^2) \,\ln N \,.
\eeq
The number of quark flavours in $b_0$ and $b_1$ is set to the number of active
flavours at the scale $\mu$, i.e.\ typically four for charm production below or near
the bottom threshold, and five for charm or bottom production
above the bottom threshold.

The presence in the resummed expressions of $\log(1-\lambda)$ gives rise to a
cut singularity starting at the branch point 
\begin{equation}
N^L_q=\exp \left(\frac{1}{b_0 \as(\mu^2)}\right) \simeq
\frac{\mu^2}{\LambdaQCD^2}\;.
\label{eq:NL}
\end{equation}
This singularity is related to the divergent behaviour of the running
coupling $\as(\mu^2)$ near the Landau pole at $\mu\simeq\LambdaQCD$, and signals the
onset of non-perturbative phenomena at very large values of $N$ or,
equivalently, when $x$ is very close to its threshold value $1$.  This
translates into an unphysical behaviour of the resummed perturbative 
result in this region.  
In Section~\ref{sec:largeN} we describe how we have dealt with this issue.

Similarly to what has been done for the quark coefficient function, in
Ref.~\cite{Cacciari:2001cw} the Sudakov-resummed expression for the initial
condition of the fragmentation function has also been derived, yielding a result
similar to Eq.~(\ref{eq:Delta_Q_S}).
To NLL accuracy we have
\begin{equation}
\label{eq:Delta_ini}
\Delta^S_{{\rm ini}}(N,\mu_{0}^2,\mQ^2) = \exp\lq \ln N \, g^{(1)}_{{\rm
  ini}}(\lambda_0) + g^{(2)}_{{\rm ini}}(\lambda_0)\rq \,,
\end{equation}
with
\beqn
\label{g1funin} 
g^{(1)}_{\rm ini}(\lambda_0) &=& 
-\frac{A^{(1)}}{2\pi b_0 \lambda_0} \;
\bigl[ 2\lambda_0 + (1-2\lambda_0) \ln (1-2\lambda_0) \bigr] \,,
\\
\label{g2funin}
g^{(2)}_{\rm ini}(\lambda_0) &=&
\frac{A^{(1)}}{2 \pi b_0}\left(\ln \frac{\mu_{0}^2}{m^2} + 2\gamma_E\right) 
\ln(1-2\lambda_0)\nn\\
&&-\frac{A^{(1)}  b_1}{4 \pi b_0^3}
\left[ 4\lambda_0 + 2 \ln (1-2\lambda_0) +  \ln^2 (1-2\lambda_0) \right]
\nn \\
&&+ \frac{1}{2\pi b_0} \left[2\lambda_0 + \ln (1-2\lambda_0) \right]
\left(\frac{A^{(2)}}{\pi b_0}\right)
+ \frac{H^{(1)}}{2 \pi b_0} \ln (1-2\lambda_0) \,,
\eeqn
and
\beq
\label{eq:lambda_0}
H^{(1)} = -\cf\,,   \quad\quad
\lambda_0 \equiv b_0 \, \as(\muz^2) \ln N\,.
\eeq
The number of quark flavours in $b_0$ and $b_1$ for the Sudakov resummation
factor of the initial condition  is set to the number of light
flavours at the scale $\mu_0 \simeq m$, i.e.\ three for charm and four for bottom.
Note that, for ease of notation, both in Eq.~(\ref{eq:Delta_Q_S})
and~(\ref{eq:Delta_ini}) the renormalization and the factorization scales
have been taken equal. The full 
expressions for the Sudakov factors can be found in~\cite{Cacciari:2001cw}.

Analogously to Eq.~(\ref{eq:Delta_Q_S}), the Sudakov-resummed part
$\Delta^S_{{\rm ini}}$ of the heavy-quark initial condition also has cut
singularities in the complex variable $N$. In the heavy-quark case the
singularities start at the branch-point 
\begin{equation}
N^L_{{\rm ini}}=\exp \left( \frac{1}{2\,b_0
\as(\muz^2)}\right) \simeq \frac{\muz}{\LambdaQCD}
\label{eq:Nini} \; ,
\end{equation}
i.e.\ at $\lambda_0=1/2$ in
Eqs.~(\ref{g1funin}) and~(\ref{g2funin}). Again, we defer to
Section~\ref{sec:largeN} the discussion of this problem.

For later convenience, we introduce here the expansions up to order $\as$ of 
$\Delta_q^S$ and $\Delta^S_{{\rm ini}}$, defined in Eqs.~(\ref{eq:Delta_Q_S})
and~(\ref{eq:Delta_ini})
\beqn
\Delta_q^S(N,q^2,\muF^2) &=& 1 + \asb(\mu^2) \,
\lq\Delta_q^S(N,q^2,\muF^2) \rq_{\as} + \ord{\as^2}\;,
\\
\Delta^S_{{\rm ini}}(N,\mu_{0}^2,\mQ^2) &=& 1 + \asb(\muz^2) \,
\lq\Delta^S_{{\rm ini}}(N,\mu_{0}^2,\mQ^2)\rq_{\as}+ \ord{\as^2}\;,
\eeqn
where
\beq
\lq\Delta_q^S(N,q^2,\muF^2)\rq_{\as} = \cf \lq  \ln^2N 
+ \(\frac{3}{2}+2\gamma_E - 2\ln\frac{q^2}{\muF^2}\) \ln N \rq \,,
\label{eq:matchCF}
\eeq
\beq
\lq\Delta^S_{{\rm ini}}(N,\mu_{0}^2,\mQ^2)\rq_{\as} =
 \cf \left[ - 2\ln^2N + 
2\left( \ln \frac{m^2}{\mu_{0}^2} - 2\gamma_E + 1 \right) \ln N  \right] \,.
\label{eq:matchINI}
\eeq
Note that they differ from the exact coefficient function
of Eq.~(\ref{eq:a_Q^1}) and from the initial condition for the fragmentation
function of Eq.~(\ref{eq:d_Q^1}) only by terms finite in the large-$N$ limit.

In order to merge the NLL-resummed and the NLO expressions without
double-counting $\ord{\as}$ logarithmic terms, we define the Sudakov-resummed
expressions for the coefficient function and initial condition in the
so-called `log-R matching scheme' as
\beqn
\label{eq:a_Q^res}
&&C_q^{{\rm res}}(N,q^2,\muF^2) =
\Delta_q^S(N,q^2,\muF^2) 
\nonumber\\
&& \quad\quad \times \exp\lg \asb(\mu^2) \lq a_q^{(1)}(N,q^2,\muF^2)
- \lq\Delta_q^S(N,q^2,\muF^2)\rq_{\as} \rq \rg \sigma_{0,q}(q^2),  
\\
\label{eq:d_Q^res}
&& D_\sQ^{{\rm res}}(N,\mu_{0}^2,\mQ^2)  =
\Delta^S_{{\rm ini}}(N,\mu_{0}^2,\mQ^2)
\nonumber\\
&& \quad\quad \times  \exp\lg \asb(\muz^2) \lq d_\sQ^{(1)}(N,\mu_{0}^2,\mQ^2) -
\lq\Delta^S_{{\rm  ini}}(N,\mu_{0}^2,\mQ^2)\rq_{\as}\rq\rg .
\eeqn
This matching prescription differs from the one employed in
Ref.~\cite{Cacciari:2001cw} (see Eqs. (36) and (76) there). However, since
the exponents in the exponentials in Eqs.~(\ref{eq:a_Q^res})
and~(\ref{eq:d_Q^res}) are small (i.e.\ do not contain large logarithms and
are of order $\as$), the exponentials can also be expanded without loss or
gain of accuracy, giving rise to different - but equivalent - matching
prescriptions, among which that of Ref.~\cite{Cacciari:2001cw}.

\subsection[The large-$N$ region]
{The large-\boldmath{$N$} region}\label{sec:largeN}
We have previously remarked how the soft-gluon resummation factors
$\Delta_q^S$ and $\Delta^S_{{\rm  ini}}$ contain singularities at large $N$
which signal the eventual failure of perturbation theory and hence 
the onset of non-perturbative phenomena.
The matching of perturbative results with non-perturbative physics is a
delicate problem, which rests, first of all, on a proper definition of 
the perturbative series. 

One way to address this problem is to work in the framework of infra-red
renormalons. In Ref.~\cite{Cacciari:2002xb}, the perturbative series is first
improved by addition of all subleading logarithms $\as^n \ln^kN$, with $k\le
n+1$, in the so-called large-$\beta_0$ approximation. The
asymptotically divergent series is subsequently regulated either by
truncation at the smallest term or with a Cauchy principal-value prescription
of its Borel antitransform. This also implicitly defines non-perturbative
terms which can be cast in the form of power corrections, hence allowing
to relate 
charm and bottom hadronization.  This procedure makes maximal use of the
insight that can be gleaned from perturbative QCD.  However, it will be shown
in Section~\ref{sec:aleph} how charm fragmentation data at $\Upsilon(4S)$
(10.6~GeV) and $Z^0$ (91.2~GeV) 
energies cannot be described simultaneously within perturbation
theory. Without a more specific understanding of the origin of this problem,
it would appear premature to even attempt to relate rigorously the charm and
bottom non-perturbative fragmentation functions.

In the present work, we do not attempt a rigorous 
formulation of the perturbative/non-perturbative matching problem and of the 
ensuing description of the non-perturbative terms.
We instead simply look for a formulation of the resummation
prescription that 
\begin{itemize}
\item[(i)] is consistent with all known perturbative results,
\item[(ii)] yields physically acceptable results,
\item[(iii)] does not introduce power corrections larger than
      generally expected for the processes in question, i.e.\ $N\Lambda/m$
      for the initial
      condition~\cite{Nason:1997pk,Jaffe:1993ie,Randall:1994gr,Cacciari:2002xb}
      and $N\Lambda^2/q^2$ for the coefficient
      functions~\cite{Dasgupta:1996ki}, where $\Lambda$ is a typical hadronic
      scale of a few hundreds MeV.
\end{itemize}

In detail, as far as the coefficient function is concerned, we make the
following replacement in Eq.~(\ref{eq:lambda}) (and
hence~(\ref{eq:Delta_Q_S}) and~(\ref{eq:matchCF})) 
\begin{equation}
 N  \to N\,\frac{1+f/N^L_q}{1+f\,N/N^L_q}\;,
\label{eq:N_tilde}
\end{equation}
where $N^L_q$ is given in Eq.~(\ref{eq:NL}) and $f$ is a parameter not
smaller than one, but of order one.  For $N \ll N^L_q$, the
replacement~(\ref{eq:N_tilde}) amounts to a tower of power corrections to
$N$, starting with $f(N-1)/N^L_q\approx f(N-1)\LambdaQCD^2/\mu^2$,
consistently with items~(i) and~(iii) listed above. Furthermore, for large
$N$, the replacement~(\ref{eq:N_tilde}) becomes $N \to N^L_q/f$.  Thus, with
this replacement, the functions $g^{(1/2)}$ of Eqs.~(\ref{g1funms})
and~(\ref{g2funms}) have no singularities in the half plane $\Re(N)>0$, so
that item~(ii) above is also fulfilled.

For the initial condition, we apply the same prescription of
Eq.~(\ref{eq:N_tilde}) in Eq.~(\ref{eq:lambda_0}) (and
hence~(\ref{eq:Delta_ini}) and~(\ref{eq:matchINI})), replacing $N^L_q$ with
$N^L_{\rm ini}$ (defined in Eq.~(\ref{eq:Nini}))
\begin{equation}
 N  \to N\,\frac{1+f/N^L_{\rm ini}}{1+f\,N/N^L_{\rm ini}}\; .
\label{eq:N_tilde_ini}
\end{equation}
In this case, the replacement 
amounts to a tower of power corrections starting with
$f(N-1)/N^L_{\rm ini}\approx f(N-1)\LambdaQCD/\mu_0$, and for large $N$
the branch cut in Eqs.~(\ref{g1funin}) and~(\ref{g2funin})
is never reached.

The Landau singularity is regulated if $f\ge 1$. For $f$ below 1 the effect
of the Landau pole should be visible. We plot in Fig.~\ref{fig:flandau-c}
the results of varying the $f$ parameter ($f=0$, i.e.\ no regulator,
$0.5$, $1$ and $1.5$), together with the pure perturbative
result, without Sudakov resummation.
\begin{figure}[htb]
\begin{center}
  \epsfig{file=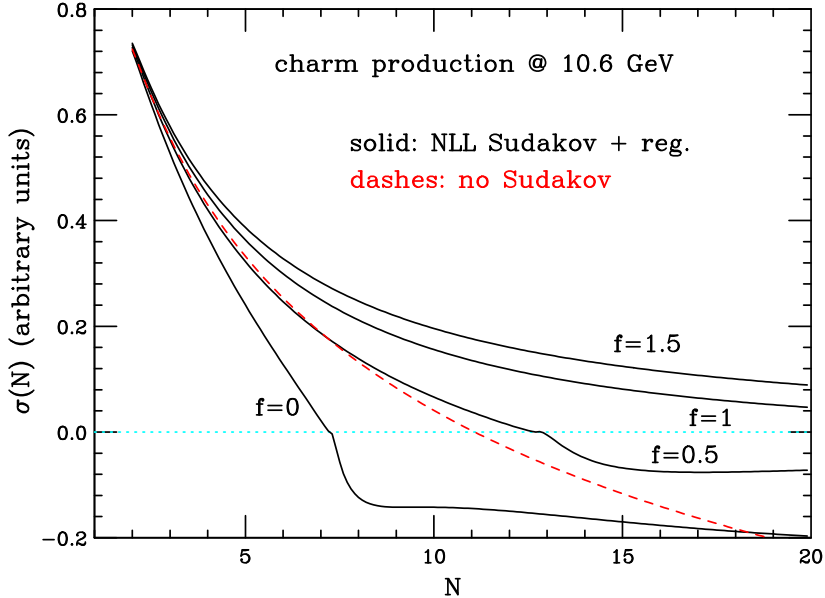,width=0.8\textwidth}
\caption{\label{fig:flandau-c} Moments of the perturbative fragmentation
function for charm production at $\sqrt{q^2}=10.6$~GeV with and without
soft-gluon resummation with different values of $f$, Eqs.~(\ref{eq:N_tilde})
and~(\ref{eq:N_tilde_ini}), in the regularization of the Landau pole
singularities.}
\end{center}
\end{figure}
First of all, we notice the tiny cusp due to the Landau singularities,
located around $N\approx 7.2$, consistently with Eq.~(\ref{eq:Nini}). The
moments become negative (and therefore unphysical) after the cusp.  With
increasing $f$, the cusp is displaced to larger values of $N$, until it
disappears for $f=1$. The fixed order cross section also changes sign at
$N\approx 11$, larger than $N^L_{\rm ini}$.  This is consistent with the
large $N$ behaviour of $d^{(1)}_Q$ shown in Eq.~(\ref{eq:d_Q^1}), such that
$D_Q(N)$ becomes negative when
\begin{equation}
N\approx \exp{\sqrt{\frac{\pi}{\cf \as(m^2)}}}\;.
\end{equation}
However, for small enough $\as$, this value should be parametrically
smaller than $N^L_{\rm ini}$. In the case of charm production, this does not
happen, reminding us that we are at the limit of validity of perturbation
theory. A more consistent behaviour is observed
in the bottom case, Fig.~\ref{fig:flandau-b}.
\begin{figure}[htb]
\begin{center}
  \epsfig{file=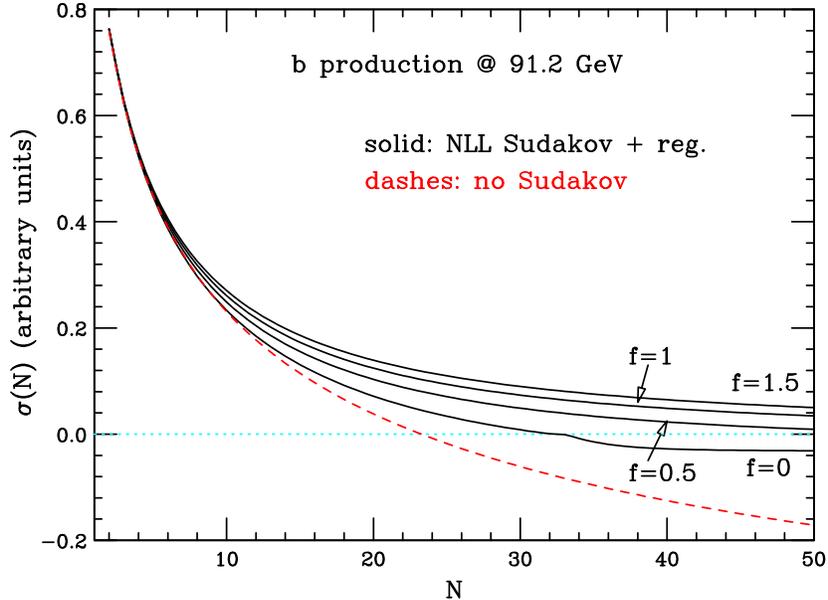,width=0.8\textwidth}
\caption{\label{fig:flandau-b}
As in Fig.~\ref{fig:flandau-c} for bottom production at
$\sqrt{q^2}=91.2$~GeV.}
\end{center}
\end{figure}
In this figure, the pattern of improvement towards the large-$N$ region,
when going from the purely-perturbative initial condition without
soft gluon resummation, to the inclusion of soft-gluon effects,
and then to the addition of the non-perturbative regularization,
is clearly visible.

We conclude our discussion with the following remarks.
We have found that, while for bottom production
the NLO result, the inclusion of Sudakov effects and
the regularization of the Landau singularities follow
numerically the correct pattern of improvements,
in the case of charm production the inclusion of
Sudakov effects induces a worse large-$N$ behaviour
of the cross section, signaling the imperfect
applicability of perturbation theory in this case.
Nevertheless, in both cases we have shown that
we can obtain a sensible physical result with
formulae that are consistent with all known results
in perturbative QCD, and modest power suppressed
effects according to the item~(iii).
We will thus apply our procedure to the fits
of charm and bottom data. We will use the value $f=1.25$
in our fits, since we found that good fits can be obtained
with this choice. We remark that the parameters of the
non-perturbative fragmentation function that we obtain
in our fits do depend upon the choice of $f$, to an extent
that can be inferred from Figs.~\ref{fig:flandau-c}
and~\ref{fig:flandau-b}. It is also clear that,
for moments around $N\approx 5$, the effect of
the inclusion of a regulated Sudakov is modest.
We will discuss in Section~\ref{sec:moment} the implications
of this fact for hadronic cross sections.

\subsection{The bottom threshold}
\label{sec:bottom_threshold}
In analogy with parton distribution functions, also parton fragmentation
functions obey matching conditions when crossing heavy-flavour thresholds.
In Ref.~\cite{Cacciari:2005ry}, we have computed these matching conditions at
next-to-leading order in the strong coupling constant $\as$ in the \MSB\
scheme. We obtain, neglecting ${\cal O}(\as^2)$ corrections
\begin{eqnarray}
D_{Q/\bar{Q}}^{(\nf)}(x,\mu_{\rm thr}^2,\mthr^2) \!\!&=&\!\! \int_x^1 \frac{dy}{y}
 \, D_g(x/y,\mu_{\rm thr}^2,\mthr^2)\,\nonumber\\
&&\times
\frac{\as}{2\pi}\,\cf\,
\frac{1+(1-y)^2}{y}
\left[\log\frac{\mu_{\rm thr}^2}{\mthr^2}-1-2\log y\right]\;
\\
 D_g^{(\nf)}(x,\mu_{\rm thr}^2,\mthr^2)\!\!&=&\!\! D_g^{(\nlf)}(x,\mu_{\rm
 thr}^2,\mthr^2) 
\left(1-\frac{\tf\as}{3\pi} \log\frac{\mu_{\rm thr}^2}{\mthr^2}\right)\;
\\
 D_{i/{\bar i}}^{(\nf)}(x,\mu_{\rm thr}^2,\mthr^2)\!\!&=&\!\! D_{i/{\bar
 i}}^{(\nlf)}(x,\mu_{\rm thr}^2,\mthr^2) 
\quad\quad\quad\mbox{for}\; i=q_1,\ldots q_\nlf \;\; ,
\end{eqnarray}
where $\nlf = \nf -1$ is the number of light flavours.
Since, in the present paper, we are interested in the evolution of
charm fragmentation function from lower scales, of the order of the charm
mass, to higher scales, these matching conditions should be used for
consistency when crossing the bottom threshold.

In this framework, at low energies (i.e.\ not much above the charm mass), the
charm is treated as a heavy quark, in order to provide a perturbative
expression for its fragmentation function.  Near the bottom threshold, the
bottom is treated as heavy, while all other quarks (including charm) are
considered light.


\subsection{Simplified evolution scheme}
For the phenomenological analysis performed in the present work,
we have numerically solved the full set of evolution equations.
It turns out, however, that, for the case of charm production
at $\Upsilon(4S)$ energies,
the contribution coming from gluon-splitting processes is fully negligible.
Our results, in this case, can thus be obtained in a simplified framework,
where only the $P_{qq}$ splitting function is kept.
The Mellin transform of $P_{qq}$
can be performed analytically, and one can work with $\nlf$ flavours,
since the annihilation energy is of the order of the bottom mass.
The evolution equation has the simple solution 
\beqn
&&E(N,\mu^2,\mu_{0}^2) =
\exp\Bigg\{
\log\frac{\as(\mu_{0}^2)}{\as(\mu^2)}\; \frac{P_{qq}^{(0)}(N)}{2\pi b_0}
\nonumber \\ \label{eq:evoqq}
&&\quad\quad\quad\quad\quad +
 \frac{ \as(\mu_{0}^2) - \as(\mu^2) }{4\pi^2 b_0}
\lq P_{qq}^{(1)}(N) - \frac{2\pi b_1}{b_0} P_{qq}^{(0)}(N) \rq
\Bigg\}\;.
\eeqn
Our final formula for the cross section, neglecting singlet contributions,
is then, using Eqs.~(\ref{eq:a_Q^res}), (\ref{eq:d_Q^res})
and~(\ref{eq:evoqq}) 
\begin{equation}\label{eq:totqq}
\sigma_\sQ(N,q^2,\mQ^2) = 
C_q^{{\rm res}}(N,q^2,\mu^2)\, E(N,\mu^2,\mu_{0}^2)
\, D_\sQ^{{\rm res}}(N,\mu_{0}^2,\mQ^2)\;.
\end{equation}
The Mellin transforms for ${a}_q^{(1)}$, $d_\sQ^{(1)}$ and $P^{(1)}_{qq}$
are given in formulae~(A.12), (A.13) and~(A.20) of
Ref.~\cite{Mele:1990cw}\footnote{We point out that there is an obvious misprint in Ref.~\cite{Mele:1990cw},
where formula (A.7) should be replaced by
\begin{eqnarray*}
\psi_m(x)=\frac{d^{m+1} \log \Gamma(x)}{d x^{m+1}}\;.
\nonumber
\end{eqnarray*}
}.

\section{Electromagnetic initial-state radiation}\label{sec:isr}
Electromagnetic initial-state radiation (ISR) can significantly affect the
single-inclusive distribution of charmed mesons. At CLEO and BELLE energies,
the hadronic cross section decreases as the inverse of the squared mass of
the hadronic system. Initial-state photon radiation is suppressed by a factor
of $\alpha_{\rm em}$, enhanced by a $\log s/m_e^2$ ($m_e$ being the electron
mass), and, depending upon how much energy is radiated away, due to the
lower hadronic mass, it is enhanced by a larger hadronic cross section. The
shape of the fragmentation function is also affected, since a consistent
fraction of events takes place at lower hadronic invariant mass.
CLEO and BELLE do not correct
their data for initial-state radiation, so, in order to perform a meaningful
fit to the fragmentation function, we have to take it into account.

We correct the initial distributions of measured inclusive cross sections bin
by bin, i.e.\ we find, by an iterative procedure, a new distribution that
reproduces the measured one after ISR has been implemented. More
specifically: calling $x_i$, $i=1,\ldots n$, the centre of the bins of the
experimental distribution, we find a distribution $D_c(x)$
(where the suffix $c$ stands for``corrected'') that is
continuous, vanishes at $x=0$ and $x=1$, and is linear in all intervals
$(0,x_1),(x_1,x_2),\ldots, (x_{n-1},x_n), (x_n,1)$, such that, when ISR
corrections are applied, we reproduce the measured distribution.

We model ISR in the following way.
We assume for the radiated electromagnetic 
energy the distribution~\cite{Kuraev:1985hb,Altarelli:1986kq,Nicrosini:1986sm,Nicrosini:1987sw}
\begin{equation}
\frac{dP}{dz} = \delta\beta(1-z)^{\beta-1} - \frac{\beta}{2}(1+z)\;,
\end{equation}
where
\begin{equation}
\beta = 2\frac{\aem}{\pi}\left[\log\frac{s}{m_e^2}-1\right]\,,\quad
\delta=1+\frac{3}{4}\beta + \frac{\aem}{\pi}\left(\frac{\pi^2}{3}-\frac{1}{2}\right)\;,\quad z=\frac{s_{\rm had}}{s}\;,
\end{equation}
$s=q^2$ being the squared centre-of-mass (CM) energy, and $s_{\rm had}$ the
square of the 
invariant 
hadronic mass. The kinematic distribution of the hadronic system is assumed
to be as if only a single photon, collinear to either the electron or the
positron, was radiated. This assumption neglects double radiation,
which gives effects of the order $\beta^2$, and the transverse momentum of the
radiated photon, which is typically much smaller than the available energy.
We use the Born cross section for the heavy-quark production in the hadronic
reference frame. 
The value of $x$ in the laboratory frame is obtained by a Lorentz
boost. 
In summary
\begin{equation}\label{eq:isrcorr}
D(x_i) = \int_{\frac{4m_h^2}{s}}^1 dz\int dy\,d \cos\theta \,
\frac{1}{\sigma_0(s)}\frac{d\sigma_0(zs,\cos\theta)}{d\cos\theta}\,
\frac{dP}{dz} D_c(y) \;\delta(x_i-x(z,y,\theta))\;.
\end{equation}
where $x(z,y,\theta)$ is the momentum fraction of the heavy flavoured hadron
(of mass $m_h$) in the $e^+e^-$ CM frame. It is obtained as follows.
We define the momentum
components of the hadron in the hadronic CM system
\begin{equation}
p_h = \frac{y}{2} \sqrt{sz-4m_h^2}\,,\quad\quad p^0_h=\sqrt{p_h^2+m_h^2},
\quad\quad p^\parallel_h=p_h\cos\theta\;,
\end{equation}
so that $y$ is its momentum fraction.
Then we boost it to
the $e^+e^-$ CM frame. Under our assumptions (that all the
electromagnetic energy is collinear either to the electron
or to the positron, and that double radiation and the
photon transverse momentum are negligible) the boost velocity is purely
longitudinal, and is given by $v=(1-z)/(1+z)$.
The hadron momentum in the $e^+e^-$ CM frame is then
\begin{equation}
p^0 = \frac{p^0_h+v p^\parallel_h}{\sqrt{1-v^2}}\,,\quad\quad p =
\sqrt{p_0^2-m_h^2}\,,
\end{equation}
and
\begin{equation}
x(z,y,\theta)=\frac{p}{\sqrt{s/4-m_h^2}}\;.
\end{equation}
We use for $\sigma_0$ the exact Born cross section in the massless
limit. Since asymmetries cancel in Eq.~(\ref{eq:isrcorr}), we always
assume the angular dependence
\begin{equation}
\frac{1}{\sigma_0(s)}\frac{d\sigma_0(zs,\cos\theta)}{d\cos\theta} =
 \frac{\sigma_0(zs)}{\sigma_0(s)}\, \frac{3}{8}\, \(1+\cos^2\theta\)\;.
\end{equation}
We do, however, supply the threshold factor to the total cross section.
Thus, near the $\Upsilon(4S)$ we have
\begin{equation}
\frac{\sigma_0(zs)}{\sigma_0(s)}=\frac{\theta(zs-4m_h^2)}{z}
\frac{ \left(1+\frac{2m^2}{sz}\right)\sqrt{1-\frac{4 m^2}{sz}}}
{ \left(1+\frac{2m^2}{s}\right)\sqrt{1-\frac{4 m^2}{s}}}\;.
\end{equation}

We checked that
the effect of finite mass corrections to the angular distribution,
the use of $m_h$ instead of the quark mass $m$ in the threshold factor, as well as
the scaling violations in the fragmentation function
due to the reduced (i.e.\ $s\to sz$) CM hadronic energy, have a negligible
impact on our results.

\begin{figure}[htb]
\begin{center}
\epsfig{file=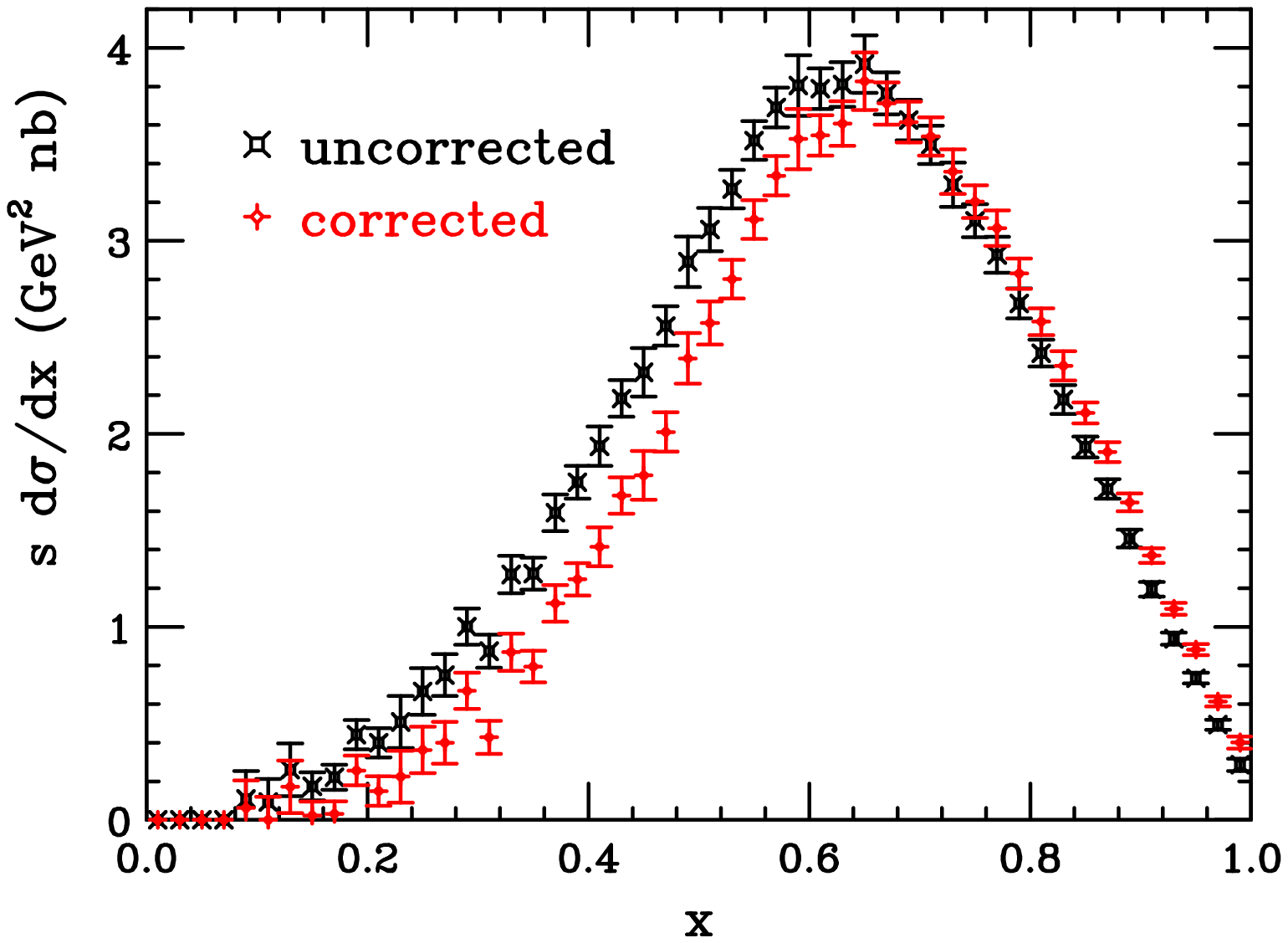,width=0.5\textwidth}~\epsfig{file=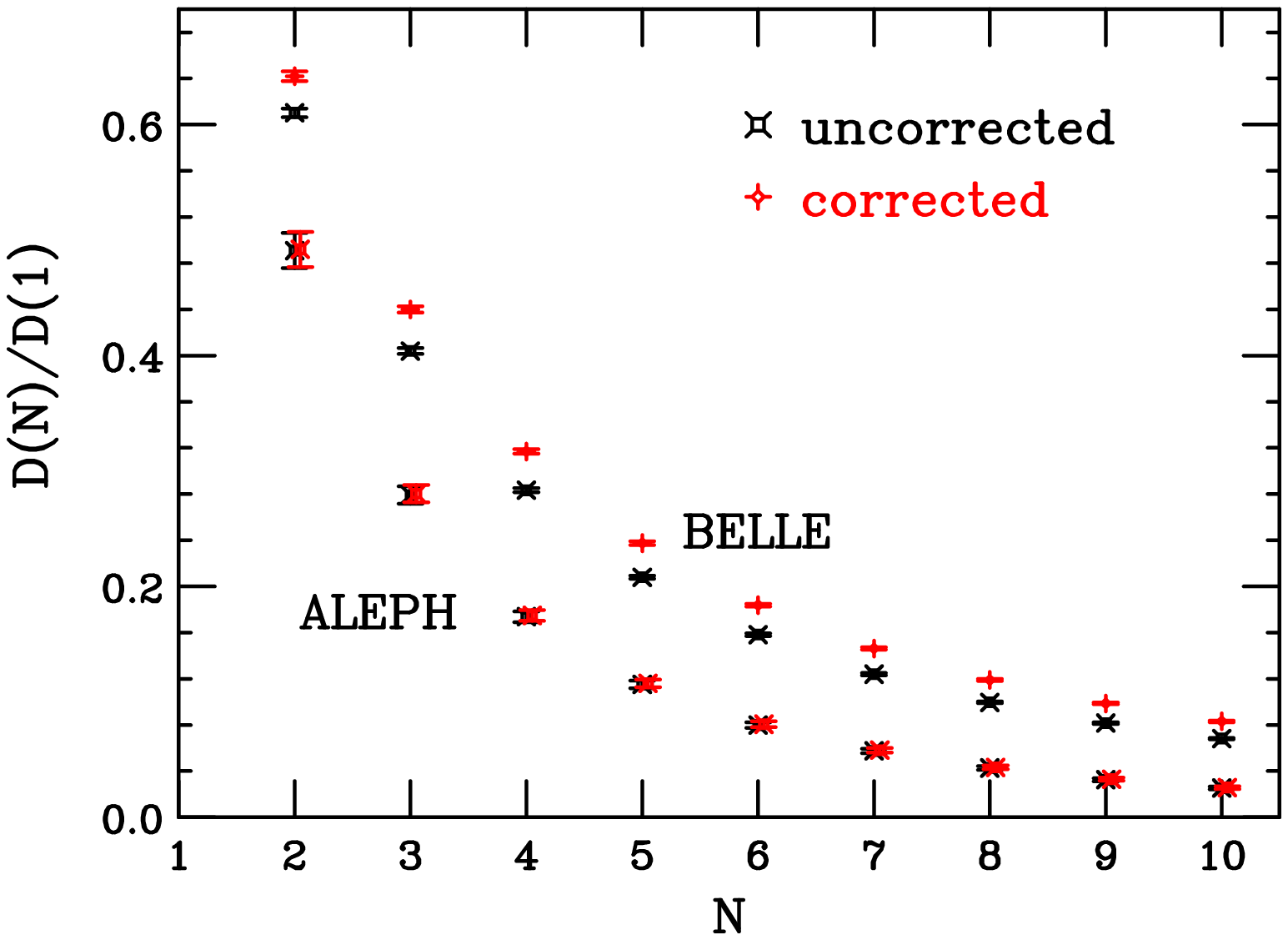,
width=0.51\textwidth}   
\caption{\label{fig:isrcorr}
Left: the effect of the ISR correction on
BELLE data for $D^{*+}\to D^{0} \pi^+$. Right: the same data in moment space,
shown together with the ALEPH ones.}
\end{center}
\end{figure}
The effect of the ISR correction is displayed in Fig.~\ref{fig:isrcorr},
where BELLE data for $D^{*+}\to D^{0} \pi^+$ are displayed before and after
the ISR correction has been applied, both in $x$ and moment space.  The ALEPH
data are also shown in moment space.  We see that the corrected spectrum for
BELLE is harder and lower in normalization than the uncorrected one. This is
to be expected, since ISR lowers the available hadronic energy, thus
softening the spectrum and increasing the cross section at the same time.  We
can also see that the effect for BELLE is not large, but nonetheless not
negligible. It is instead much less prominent, up to the point of being
negligible, for the ALEPH data taken on the $Z^0$ peak, as expected.

\section{Non-perturbative fragmentation function}\label{sec:nonpFF}
In the heavy-quark fragmentation-function formalism, the largest
non-perturbative effects come  
from the initial condition, since one expects power corrections
of the form $\Lambda/\mQ$.
We assume that all these
effects can be described by a non-perturbative fragmentation function
$\Dnp$, that takes into account all low-energy effects, including
the process of the heavy quark turning into a
heavy-flavoured hadron, that has to be convoluted with the perturbative cross
section.
Thus, the Mellin transform of the
full resummed cross section, including
non-perturbative corrections, is
\beq
\label{eq:hadfactor}
   \sigma_{\sss H}(N,q^2) = \sigma_\sQ(N,q^2,\mQ^2)  \Dnp (N) \;.
\eeq
We have attempted to fit CLEO and BELLE $D^*$ data using several forms for
$\Dnp$.
We found that the best fits are obtained with the two-component form
\begin{equation}
\Dnp(x)={\rm Norm.} \times \frac{1}{1+c}
\left[ \delta(1-x) + c N_{a,b}^{-1} (1-x)^a x^b\right]\;,
\label{eq:threepar}
\end{equation}
with
\begin{equation}
N_{a,b}=\int_0^1 (1-x)^a x^b\;.
\end{equation}
This form is a superposition of a maximally hard component (i.e.\ the delta function)
and the form proposed in Ref.~\cite{Colangelo:1992kh}.
It can be given a simple phenomenological interpretation,
the hard term corresponding in some sense to the direct exclusive
production of the $D^*$, and the Colangelo-Nason form accounting for
$D^*$'s produced in the decay chain of higher resonances.

Following
the approach  of Ref.~\cite{Cacciari:2003zu}, we assume that the $D$ meson 
non-perturbative fragmentation function is the sum of a direct component, which
is isospin invariant, plus the component arising from the $D^*$ decay.
The decay $D^*\to D\pi$ is very close to threshold, so that the $D$ has the same
velocity of the $D^*$, and their momenta are thus proportional to their masses.
Under these circumstances, the component of the $D$ fragmentation function
arising from $D^* \to D\pi$ decays is given by
\begin{equation}
B(D^*\to D\pi)\; \pDnp^{D}(x)\;,
\end{equation}
where we have defined
\begin{equation}
\pDnp^{D}(x) =  \Dnp^{D^{*}}\(x \frac{m_{D^*}}{m_{D}}\)\;\frac{m_{D^*}}{m_{D}}
\;\theta\(1-x \frac{m_{D^*}}{m_{D}}\)\,,
\end{equation}
and $B(D^*\to D\pi)$ is the branching ratio of $D^* \to D\pi$.
Observe that $\pDnp^{D}$ has been defined so as to have the same normalization
as $\Dnp^{D^{*}}$. In $N$ space we obtain immediately
\begin{equation}
\pDnp^{D}(N) =  \Dnp^{D^{*}}(N) \left[\frac{m_D}{m_{D^*}}\right]^{N-1}\;.
\end{equation}

For the $D^*\to D\gamma$ decay, in the $D^*$ frame,
the $D$ has non-negligible velocity,
but it is non-relativistic. We call $\theta$ its decay
angle with respect to the $D^*$ direction, and 
$p_D$ its momentum
\begin{equation}
p_D=\frac{m_{D^*}^2-m_D^2}{2m_{D^*}}\;.
\end{equation}
We call $\beta$ the $D^*$ velocity and $\gamma=1/\sqrt{1-\beta^2}$.
Thus, the longitudinal component of
the $D$ momentum in the laboratory frame is given by
a Lorentz boost
\begin{equation}
\gamma \( p_D\cos\theta+\beta m_D \)\;,
\end{equation}
where we have neglected terms of order $p_D^2$.
Thus the component of the $D$ fragmentation function
coming from $D^*\to D\gamma$ decay is given by
\begin{equation}
B(D^*\to D\gamma) \; \gDnp^D(x)\;,
\end{equation}
with
\begin{equation}
\gDnp^{D}(x)=\int dy \,\frac{d\cos\theta}{2}\, \Dnp^{D^*}(y)\; \delta\left(
\frac{\gamma(p_D\cos\theta+\beta m_D)}{p_{\rm max}}-x\right) \,,
\end{equation}
where $p_{\rm max}$ is the maximum $D$ momentum in the laboratory.
Since we always consider the ultra relativistic limit, we have
\begin{equation}
y=\frac{\gamma m_{D^*}}{p_{\rm max}}\;,\quad \beta \to 1 \;,
\end{equation}
so that we obtain
\begin{equation}
\gDnp^{D}(x)=\int dy\, \frac{d\cos\theta}{2}\, \Dnp^{D^*}(y)\; \delta\left(
\frac{y(p_D\cos\theta+m_D)}{m_{D^*}}-x\right) \,.
\end{equation}
The double integral cannot be performed in closed form.
However, it is easy to obtain the moments
\begin{eqnarray}
\gDnp^{D}(N) &=&  \Dnp^{D^*}(N) \int \frac{d\cos\theta}{2} \left[\frac{p_D\cos\theta+m_D}{m_{D^*}}\right]^{N-1}
\nonumber \\
&=& \Dnp^{D^*}(N)\, \frac{m_{D^*}}{2p_D}\, \frac{(m_D+p_D)^N-(m_D-p_D)^N}{N m_{D^*}^N}\;.
\end{eqnarray} 
We thus describe $D^{+/0}$ production as the sum of a primary (i.e.\ not coming
from $D^*$ decays) component, plus the contributions coming from
$D^*$ decays
\begin{eqnarray}
\Dnp^{D^{+}}(x) &=& \Dnp^{D^+,p}(x)
 + B(D^{*+}\to D^{+}\pi^0) \pDnp^{D^{+}}(x)
\nonumber \\&&
 + B(D^{*+}\to D^{+}\gamma) \gDnp^{D^{+}}D(x)\;,
\\
\Dnp^{D^0}(x) &=& \Dnp^{D^0,p}(x)
 + [B(D^{*+}\to D^0\pi^+)+ B(D^{*0}\to D^{0}\pi^0)] \pDnp^{D^{0}}(x)
\nonumber \\ &&
+B(D^{*0}\to D^{0}\gamma) \gDnp^{D^{0}}(x)\;.
\end{eqnarray}
We took the value of masses and branching ratios from
Ref.~\cite{Eidelman:2004wy}. 
For reference, we report in Table~\ref{tab:decrates} the values we used for
the masses and for the decay rates of the charmed mesons.
\begin{table}[htb]
\begin{center}
\begin{tabular}{|l|c|}
\hline
 $D$ (mass in~GeV)& branching ratios  \\
\hline
$         D^{*0}(2006.7 \pm 0.4) \to D^0 \pi^0$   & $0.619 \pm 0.029$\\
$\phantom{D^{*0}(2006.7 \pm 0.4)}\to D^0 \gamma$  & $0.381 \pm 0.029$\\
\hline
$         D^{*+}(2010.0 \pm 0.4) \to D^0 \pi^+$   & $0.677 \pm 0.005$\\
$\phantom{D^{*+}(2010.0 \pm 0.4)}\to D^+ \pi^0$   & $0.307 \pm 0.005$\\
$\phantom{D^{*+}(2010.0 \pm 0.4)}\to D^+ \gamma$  & $0.016 \pm 0.004$ \\
\hline
$D^{0}(1864.5 \pm 0.4) \to K^- \pi^+$             & $0.0381 \pm 0.0009$\\
$D^{+}(1869.3 \pm 0.4) \to K^- \pi^+ \pi^+ $      & $0.092  \pm 0.006$\\
\hline
\end{tabular}
\end{center}
\caption{\label{tab:decrates} Charm hadron masses and branching ratios.}
\end{table}



\section{\boldmath{$D$} mesons data fits near the \boldmath{$\Upsilon(4S)$}}
\label{sec:cleobelle}
Several parameters enter our calculations. First of all, at all matching
points, there are scale choices that could be varied, to yield a perturbative
uncertainty in our result. Those are the initial evolution scale $\mu_0$, the
matching scale for the crossing of the $b$ threshold $\mu_{\rm thr}$, and the
final evolution scale $\mu$.  In the present work we fix
\begin{equation}
\mu_0=m\,,\quad \mu=\sqrt{q^2}\,,\quad \mu_{\rm thr}=\mthr=m_b\;.
\end{equation}
These scales could, in principle, be varied by a factor of order two
around the values listed above, yielding a sensibly different
result. However, in general, the scale variation
will simply result in different values for the fitted parameters
of the non-perturbative form. When computing cross sections
for different processes, one should then use the parametrization
appropriate for the scale choice that has been made in the fit, hence
compensating for the change.
In the present work we will not pursue this issue further,
since our aim is simply to show that a fit within QCD is possible.
A similar remark applies to the value of $\LambdaQCD$ and the quark masses,
that we will fix at
\begin{equation}
\Lambda^{(5)}_{\rm QCD}=0.226\;\mbox{GeV}\,,\quad m_c=1.5 \;\mbox{GeV}\,,
\quad m_b=4.75\;\mbox{GeV}\,.
\end{equation}
The CLEO and BELLE data are given as absolute cross sections.
Since we correct the data for ISR effects, we should
normalized our data to the $e^+e^-$ charm cross section
corrected for ISR effects. We thus use the value of $R(e^+e^-)$
reported in Ref.~\cite{Ammar:1997sk}, defining
\begin{equation}
\sigma_c(s) =
\sigma^{(0)}_{\mu^+\mu^-}(s)\times 3.56 \times 0.4 \times  2\;,
\end{equation}
where $\sigma^{(0)}_{\mu^+\mu^-}(s)=86.86\,{\rm nb}\,/s$ is the
Born cross section for $e^+e^- \to \mu^+\mu^-$, $3.56$ is the
value of $R$ measured by CLEO, 0.4 is the charm fraction, and the factor
of 2 allows for the counting of both charge conjugate states.

We have fitted all $D^{*+}$ and $D^{*0}$ data with the same set
of parameters, except for the normalization, which is kept
independent for each data set. This procedure is justified,
since the errors in the data do not include overall errors
that do not affect the shape of the fragmentation function.
We have limited ourselves to the fit range $ 0.2<x<1$ for CLEO
and $0.08<x<1$ for BELLE. In the case of BELLE data, we use only the
continuum sample for $x<0.5$, 
and for $x>0.5$ we combine the continuum and the on-resonance sample
in the following way
\begin{equation}
y=\frac{y_c/s_c^2+y_r/s_r^2}{1/s_c^2+1/s_r^2}\,,\quad
s=\frac{1}{\sqrt{1/s_c^2+1/s_r^2}}\,,\quad
d=\frac{d_c/s_c^2+d_r/s_r^2}{1/s_c^2+1/s_r^2}\;,
\end{equation}
where $y_{c/r}$, $s_{c/r}$ and $d_{c/r}$ are the central value,
the statistical error and the systematic error of the
continuum/on-resonance data, and $y$, $s$ and $d$ are our combined
central value, statistical error and systematic error.
For all data sets we combine the statistical and systematic errors
in quadrature.

\begin{table}[htb]
\begin{center}
\begin{tabular}{|l|c|c|c|c|c|}
\hline
\multicolumn{6}{|c|} 
{Eq.~(\protect{\ref{eq:threepar}}): $a=1.8 \pm 0.2$, \ $b=11.3 \pm 0.6$, \ 
  $c=2.46\pm 0.07$, \ total  $\chi^2= 139 $ } 
\\ \hline
 Set &
(C) $D^{*+}$ & (B) $D^{*+}\to D^0$ & (B) $D^{*+}\to D^+$ &
(C) $D^{*0}$ & (B)  $D^{*0}$
\\ \hline
Norm. & $0.238$ & $0.253$ & $0.227$ &$0.225$ & $0.211$
\\ \hline 
$\chi^2/$pts & 33/16 & 63/46 & 13/46& 13/16 & 17/46
\\ \hline
\end{tabular}
\caption{\label{tab:fitdstar}
Results of the fit to $D^*$ CLEO (C) and BELLE (B) data.
The last line reports the $\chi^2$ over the number of fitted points
for each data set.}
\end{center}
\end{table}
The result of the fit is reported in Table~\ref{tab:fitdstar} and
in Figs.~\ref{fig:CLEOdstarp}, \ref{fig:BELLEdstarptoD0},
\ref{fig:BELLEdstarptoDp}, \ref{fig:CLEOdstar0} and~\ref{fig:BELLEdstar0}
we show the data and the fitted curve, both in $x$ and moment space.
\begin{figure}[htb]
\begin{center}
  \epsfig{file=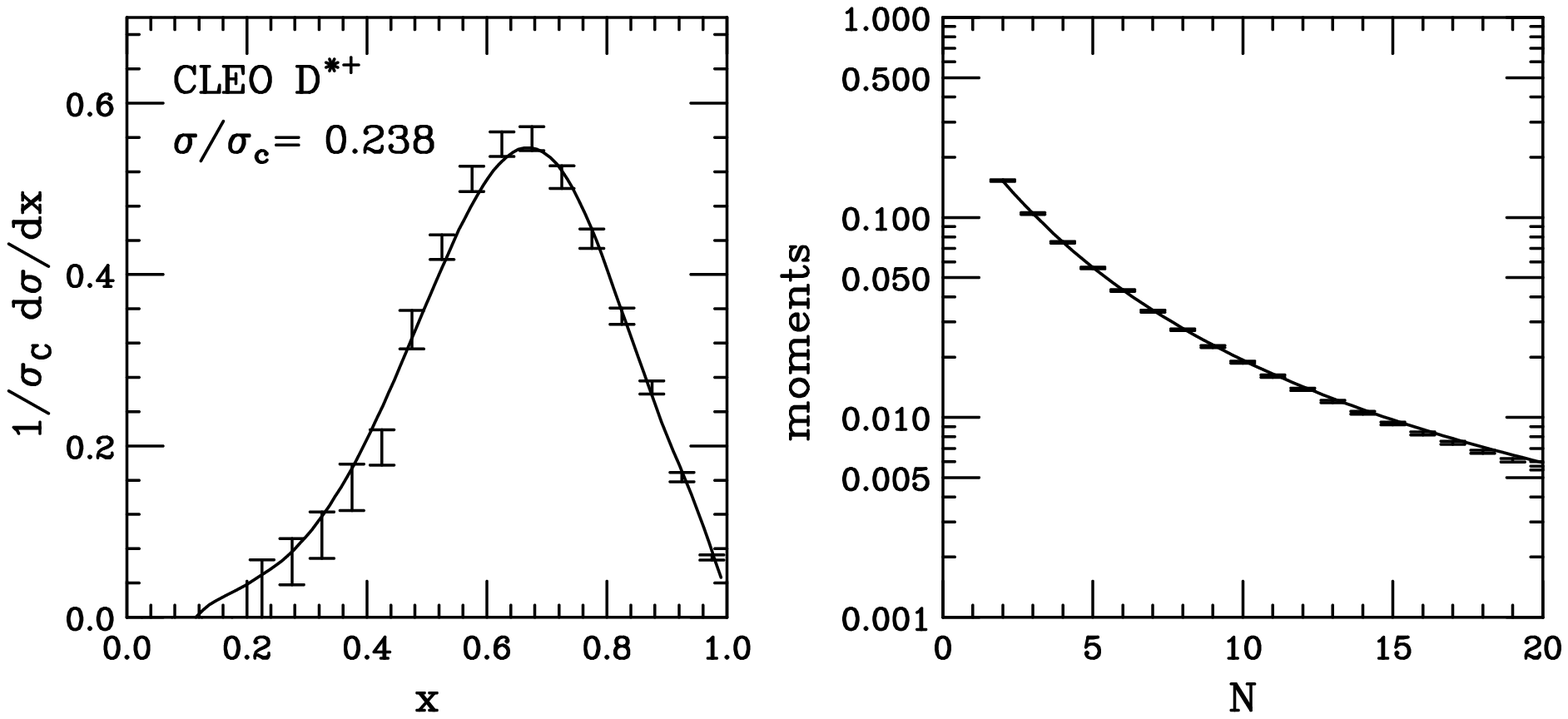,width=\textwidth}
\caption{\label{fig:CLEOdstarp}
Fit to CLEO $D^{*+}$ data.}
\end{center}
\end{figure}
\begin{figure}[htb]
\begin{center}
  \epsfig{file=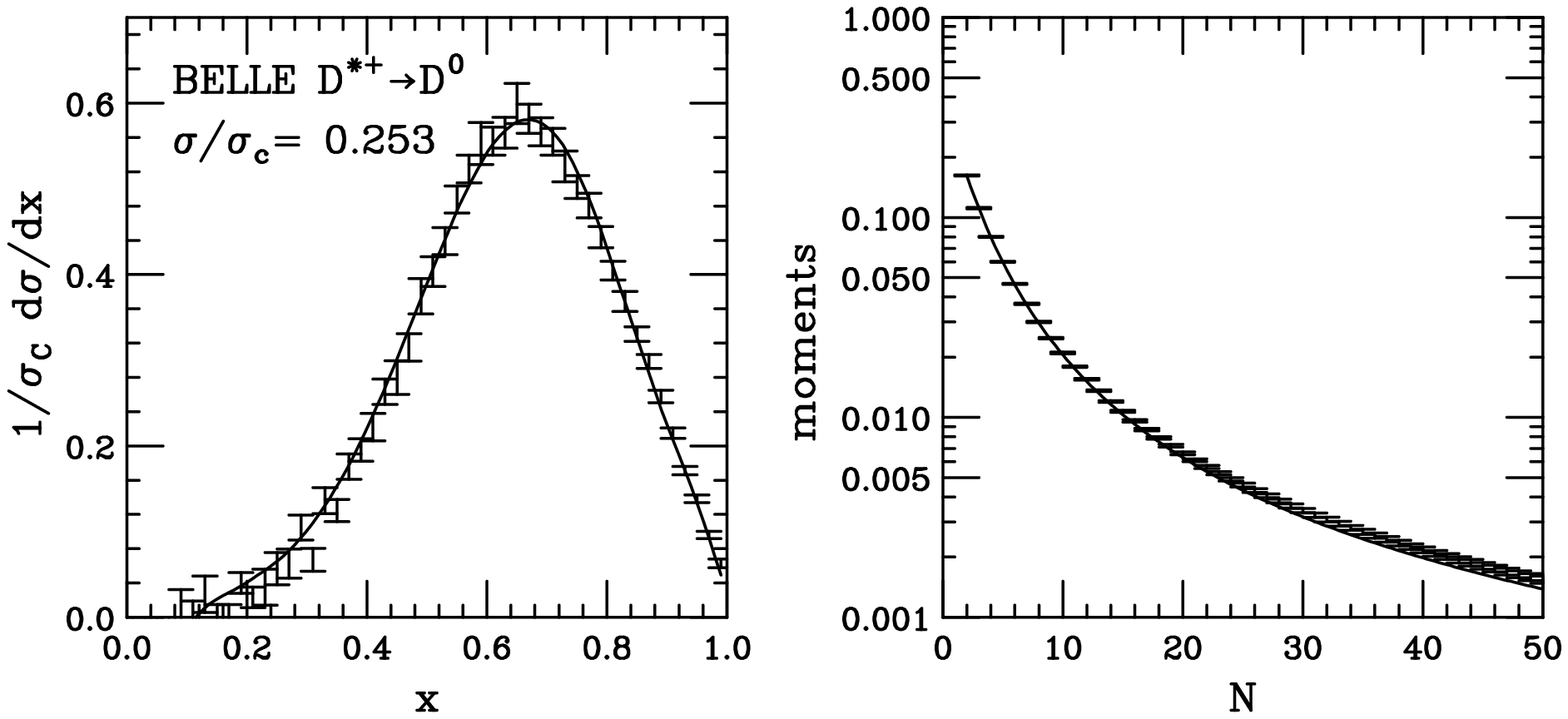,width=\textwidth}
\caption{\label{fig:BELLEdstarptoD0}
Fit to BELLE $D^{*+}\to D^0$ data.}
\end{center}
\end{figure}
\begin{figure}[htb]
\begin{center}
  \epsfig{file=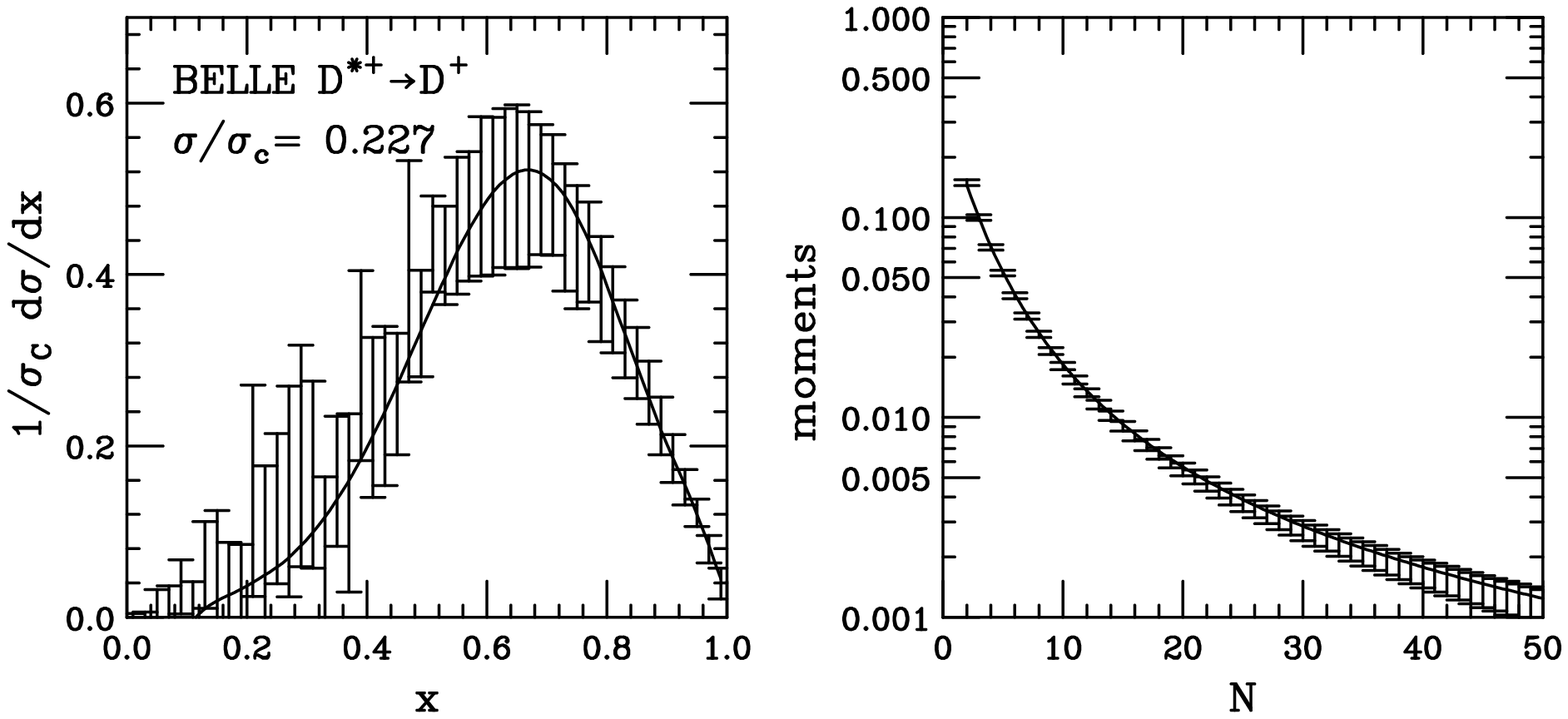,width=\textwidth}
\caption{\label{fig:BELLEdstarptoDp}
Fit to BELLE $D^{*+}\to D^+$ data.}
\end{center}
\end{figure}
\begin{figure}[htb]
\begin{center}
  \epsfig{file=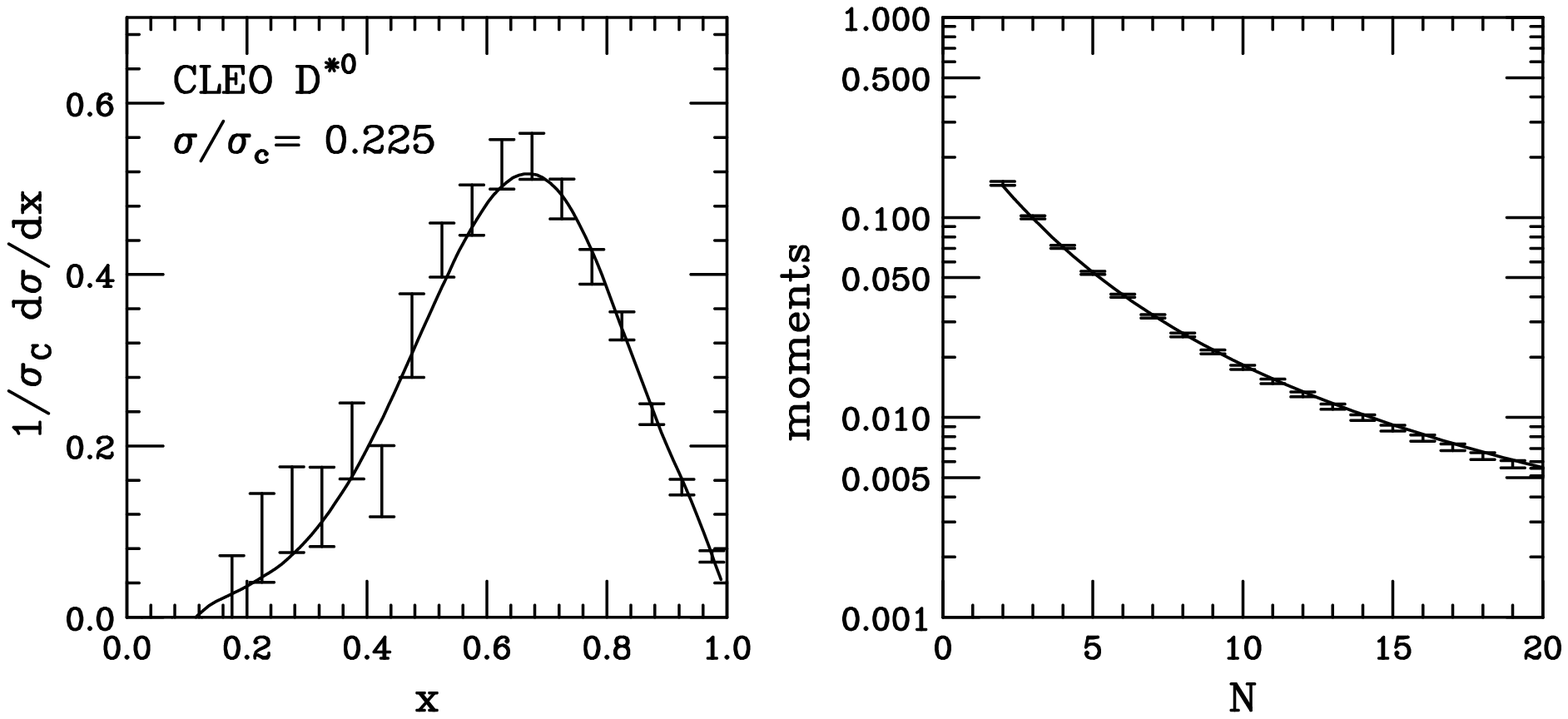,width=\textwidth}
\caption{\label{fig:CLEOdstar0}
Fit to CLEO $D^{*0}$ data.}
\end{center}
\end{figure}
\begin{figure}[htb]
\begin{center}
  \epsfig{file=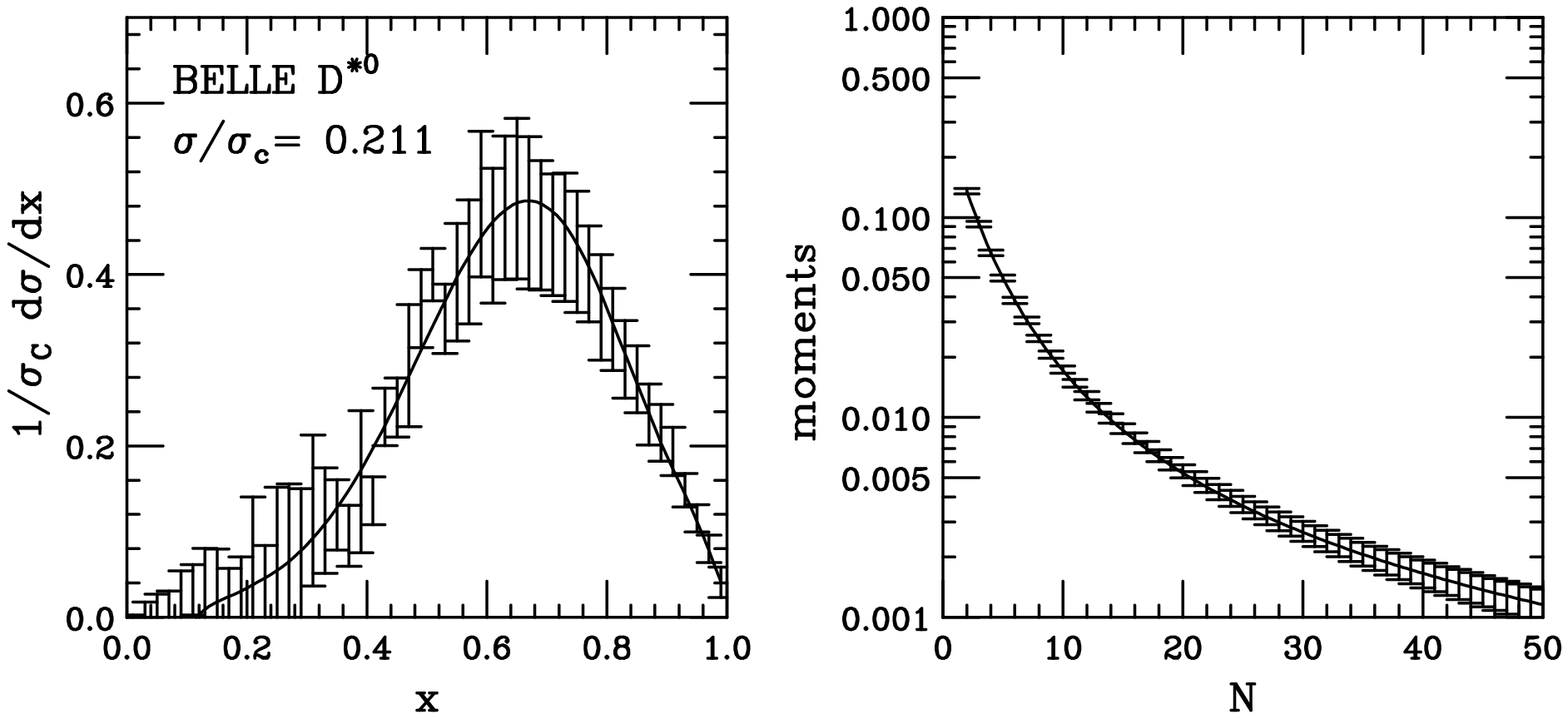,width=\textwidth}
\caption{\label{fig:BELLEdstar0}
Fit to BELLE $D^{*0}$ data.}
\end{center}
\end{figure}

A considerable part of $D$'s are produced indirectly through
$D^*$ decays. Here we assume that both $D^*$'s and the $D$'s
that are not the product of $D^*$ decay are produced
with a charge-independent rate. Under this assumption,
the fraction of ``direct'' (meaning not arising from $D^*$)
and ``indirect'' (from $D^*$) $D$ mesons are
in relative proportion of $0.473$ to $0.527$. These numbers
can be extracted from the total production cross section
of charmed mesons reported in Ref.~\cite{Artuso:2004pj}, and from
Table IX of Ref.~\cite{Seuster:2005tr}. 
We then use the parametrization of Table~\ref{tab:fitdstar} for the $D^*$
production, the branching ratios for $D^*\to D$ decays given in
Table~\ref{tab:decrates} and a description of the decay as detailed in
Sec.~\ref{sec:nonpFF}.

We parametrize the direct $D$ component with the same form used for
the $D^*$, and fit it to the $D^+$ production data, where
a larger fraction of direct $D$ is expected.
We then use the fitted direct $D$ parametrization to describe the
direct part of the $D^0$ production data. In all cases, the overall
normalization is chosen for a best fit to each data set, in order
to be insensitive to overall normalization differences.

The result of the fit for the $D^{+/0}$ mesons
is reported in Table~\ref{tab:fitd}.
\begin{table}[htb]
\begin{center}
\begin{tabular}{|l|c|c|c|c|}
\hline
\multicolumn{5}{|c|} 
{Eq.~(\protect{\ref{eq:threepar}}):  $a=1.1 \pm 0.1$, \ $b=7.6 \pm 0.6$, \
  $c=4.6\pm 0.2$ 
} 
\\ \hline
& \multicolumn{2}{|c|}{total $\chi^2=50$} &   
  \multicolumn{2}{|c|}{total $\chi^2=109$} \\
\hline
 Set &
(C) $D^{+}$ & (B) $D^{+}$ & (C) $D^0$ & (B) $D^0$
\\ \hline
Norm. & $0.263$ & $0.270$ & $0.609$ &$0.598$
\\ \hline 
$\chi^2/$pts & 14/16 & 36/46 & 32/16 & 77/46
\\ \hline
\end{tabular}
\caption{\label{tab:fitd}
Results of the fit to $D$ CLEO (C) and BELLE (B) data.
The fit was performed over the $D^+$ data only, that are
more sensitive to the direct component, and then used to
describe $D^0$ data.}
\end{center}
\end{table}
In Figs.~\ref{fig:CLEOdp}, \ref{fig:BELLEdp},
\ref{fig:CLEOd0} and~\ref{fig:BELLEd0}
we show the data and the fitted curve, both in $x$ and moment space.
\begin{figure}[htb]
\begin{center}
  \epsfig{file=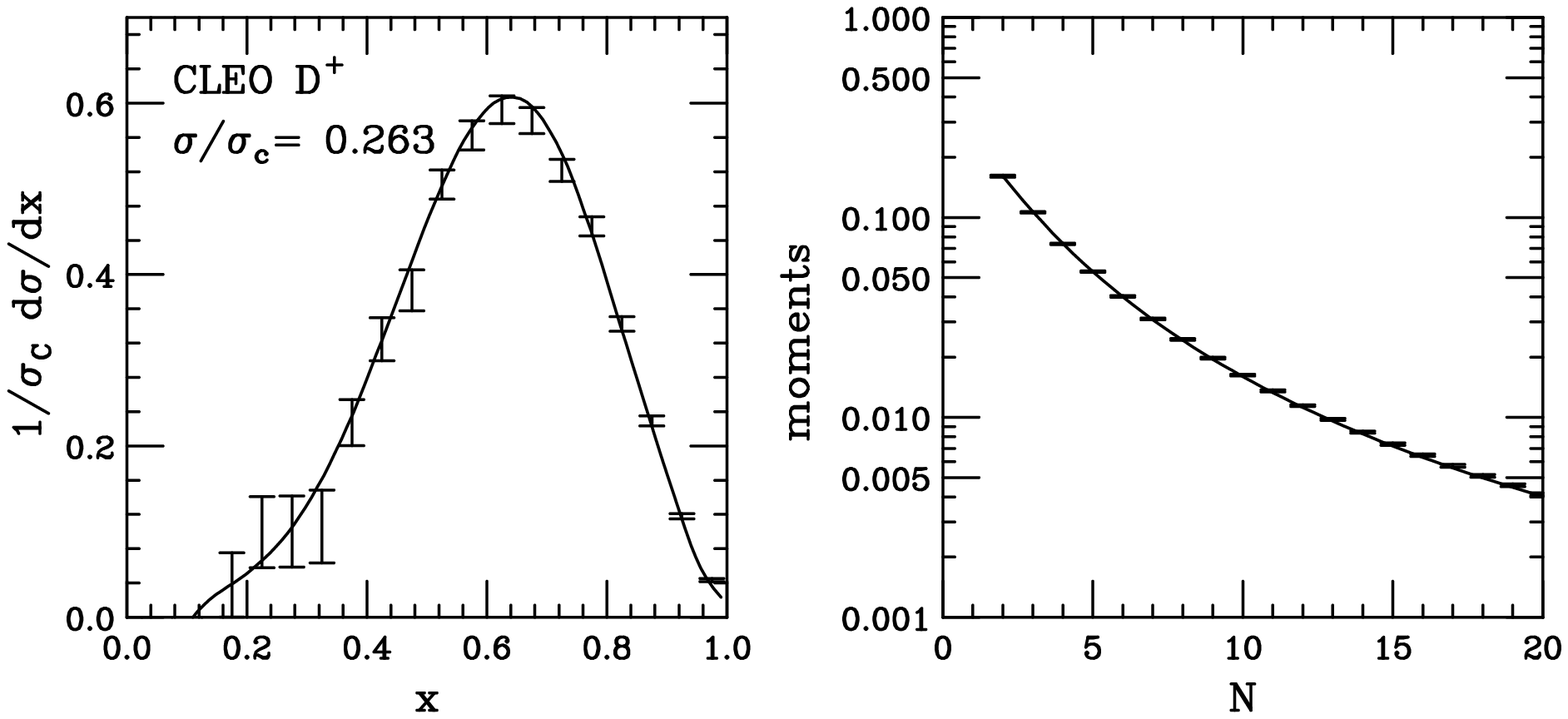,width=\textwidth}
\caption{\label{fig:CLEOdp}
Fit to CLEO $D^{+}$ data.}
\end{center}
\end{figure}
\begin{figure}[htb]
\begin{center}
  \epsfig{file=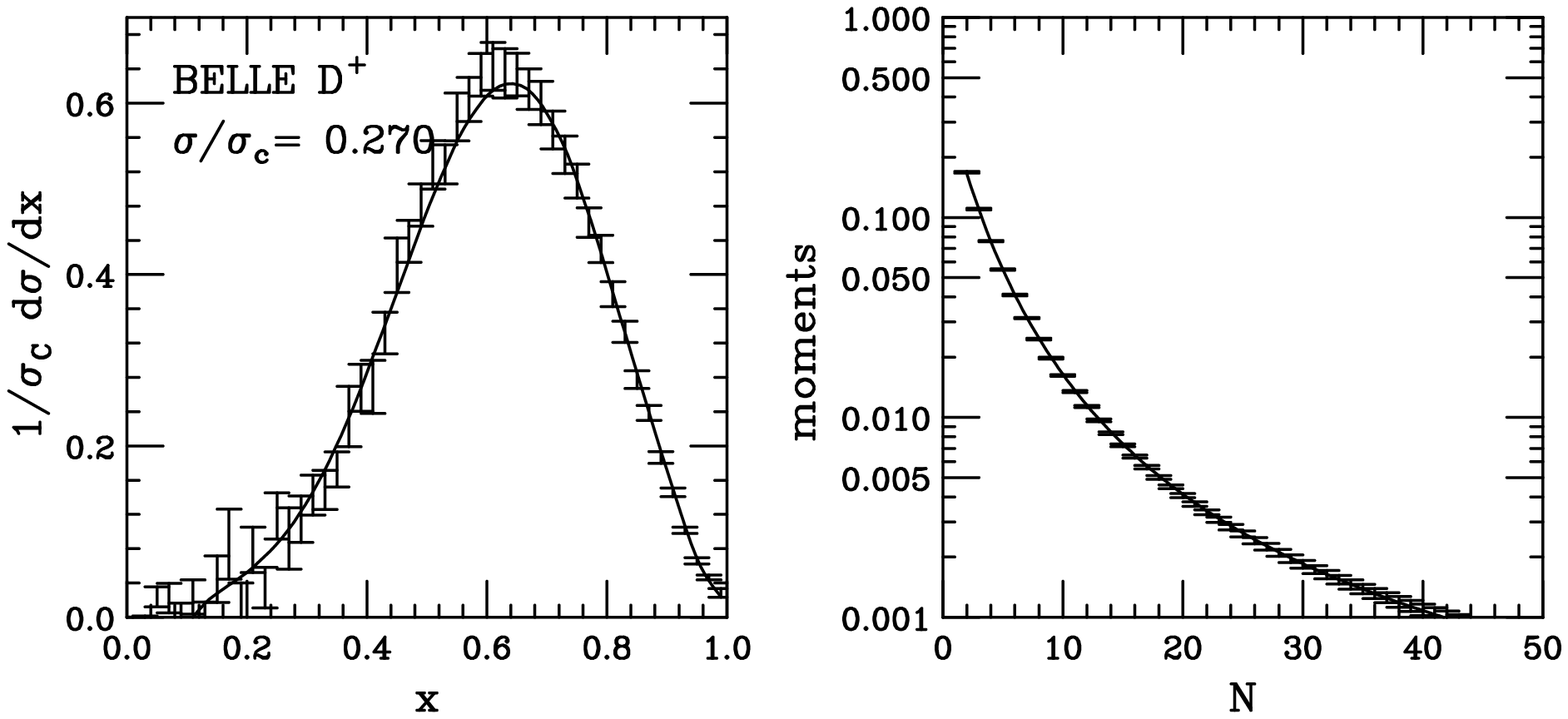,width=\textwidth}
\caption{\label{fig:BELLEdp}
Fit to BELLE $D^{+}$ data.}
\end{center}
\end{figure}
\begin{figure}[htb]
\begin{center}
  \epsfig{file=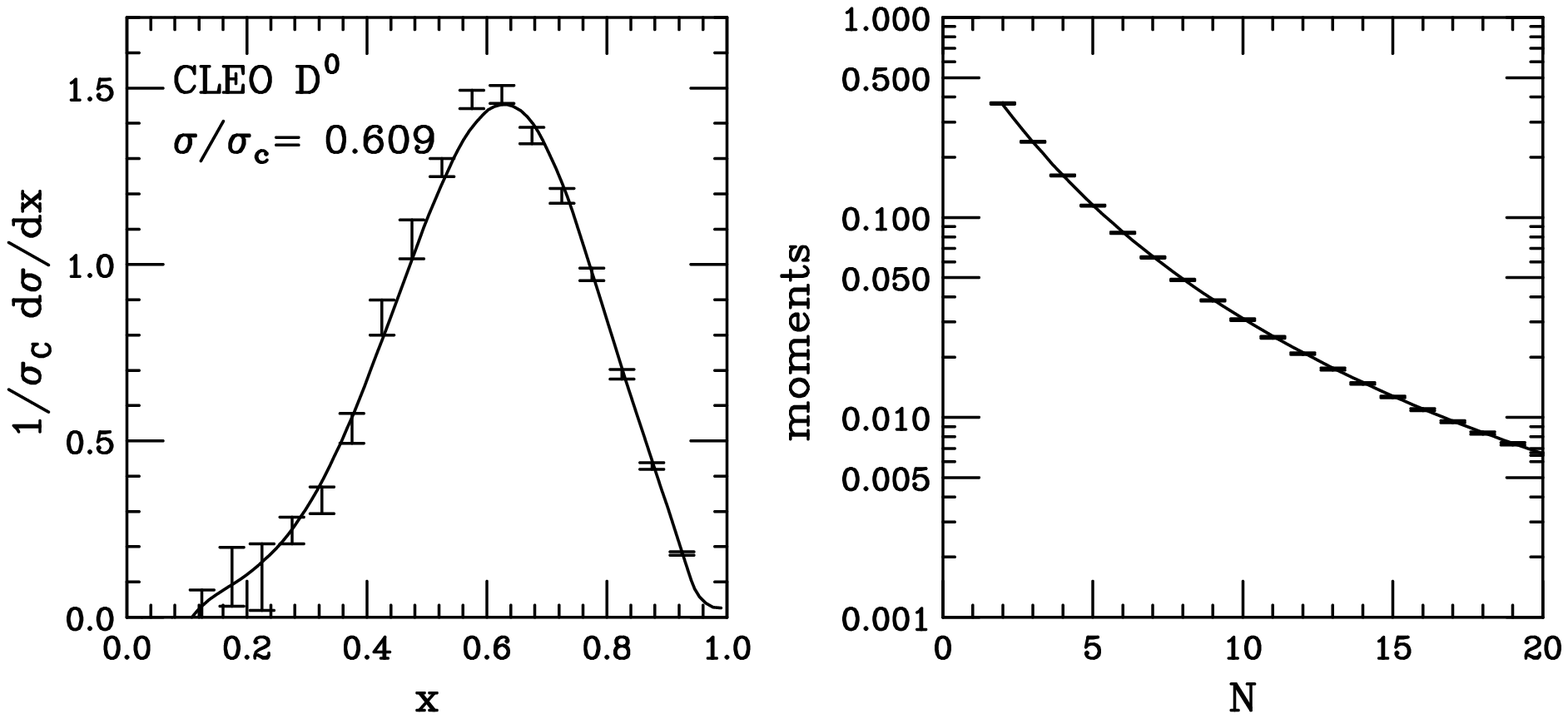,width=\textwidth}
\caption{\label{fig:CLEOd0}
CLEO $D^{0}$ data and the best fit extracted from $D^+$ data.}
\end{center}
\end{figure}
\begin{figure}[htb]
\begin{center}
  \epsfig{file=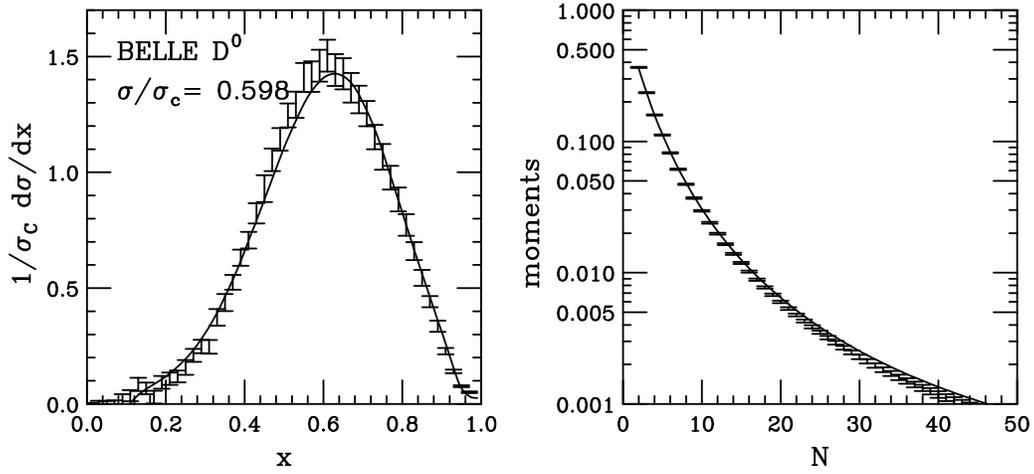,width=\textwidth}
\caption{\label{fig:BELLEd0}
BELLE $D^{0}$ data and the best fit extracted from $D^+$ data.}
\end{center}
\end{figure}
\clearpage

\section{\boldmath{$D$} mesons data fits on the \boldmath{$Z^0$}}
\label{sec:aleph}
In this work we use the
data from the ALEPH collaboration~\cite{Barate:1999bg},
which are the most precise ones. These data are
affected by electromagnetic
initial-state radiation as well. However, unlike the
$\Upsilon(4S)$ case, the ISR does not appreciably distort the spectrum,
but it mostly affects the total cross section.
This is easily understood: any appreciable
amount of ISR
on the $Z^0$ peak brings the reaction off
resonance, to a vanishing cross section.
Thus, the bulk of heavy flavour production always
takes place on the $Z^0$ peak. Conversely, at the $\Upsilon(4S)$
energy, the ISR generates processes with higher cross section
and a lower hadronic invariant mass.
Since the ALEPH data are normalized to the total number
of hadronic events, the effects of ISR largely cancel in the ratio.
We shall anyway
perform the correction for ISR for these data as well.

In Fig.~\ref{fig:ALEPH-Dstarp-fit} we display our fit with ALEPH data.
\begin{figure}[htb]
\begin{center}
  \epsfig{file=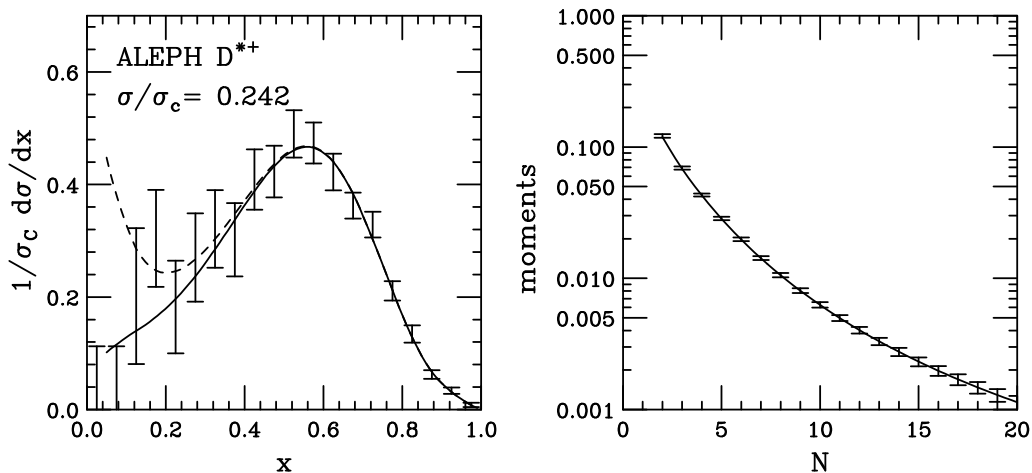,width=\textwidth}
\caption{\label{fig:ALEPH-Dstarp-fit}
ALEPH $D^{*+}$ data and the result of our non-singlet fit (solid line).
The dashed line represents the result obtained using full evolution.}
\end{center}
\end{figure}
We fit the data in the region $x \in [0.4,1]$ using the non-singlet component
only, since a subtraction of the gluon-splitting contributions was performed
by ALEPH.  Observe that, in this calculation, the bottom-threshold crossing
has to be dealt with, according to the discussion of
Section~\ref{sec:bottom_threshold}.  We also show, for comparison, the full
evolution result (dashed line), using the same parameters obtained in the
non-singlet fit.  As we can see, the difference is only visible at small $x$.
The result of the fit for the non-perturbative parameters is
\begin{equation}\label{eq:ALEPHDstarfit}
a=2.4\pm 1.2\,,\quad  b=13.9\pm 5.7\,\quad c=5.9\pm 1.7\,,
\end{equation}
with a $\chi^2=4.2$ for 13 fitted points. These results are
not really consistent with
those for the $\Upsilon(4S)$ data in Tab.~\ref{tab:fitdstar}.

In order to better quantify the discrepancy between
Eq.~(\ref{eq:ALEPHDstarfit}) and Tab.~\ref{tab:fitdstar}
we use the parametrization of CLEO and BELLE data
to predict the $D^{*}$ fragmentation
function at LEP energies.
The LEP prediction, using the parametrization
of Table~\ref{tab:fitdstar}, is reported in
Fig.~\ref{fig:ALEPHDstarp} together with ALEPH data.
\begin{figure}[htb]
\begin{center}
  \epsfig{file=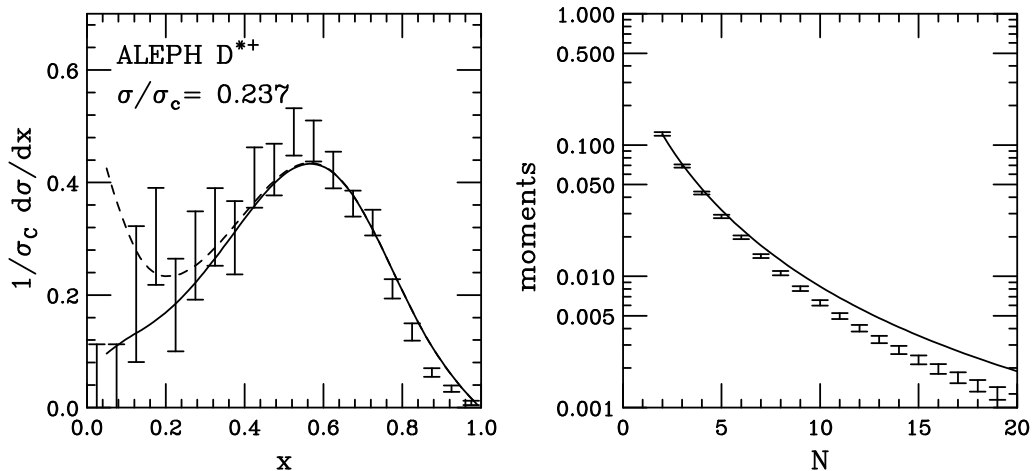,width=\textwidth}
\caption{\label{fig:ALEPHDstarp}
ALEPH $D^{*+}$ data, compared to the QCD prediction.}
\end{center}
\end{figure}
We observe that the fitted normalization is very close
to the CLEO $D^{*+}$ normalization.
We find a $\chi^2=60.1$ (for 13 fitted points) for this parametrization.
Thus, the description is not satisfactory, especially in the
large-$x$ (large-$N$) region.

In Fig.~\ref{fig:ALEPHoTH} we show the ratio of the moments
of ALEPH $D^{*+}$ data over our prediction.
\begin{figure}[htb]
\begin{center}
  \epsfig{file=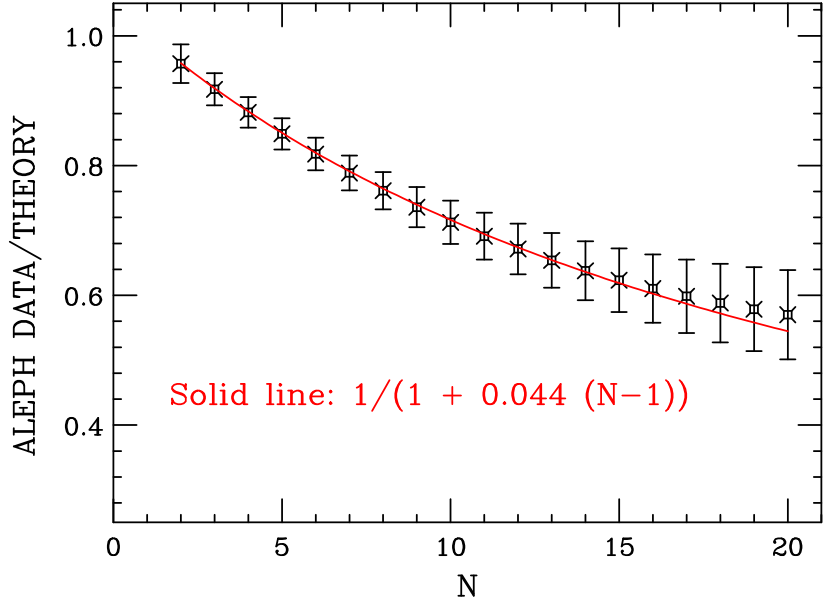,width=0.8\textwidth}
\caption{\label{fig:ALEPHoTH}
ALEPH $D^{*+}$ data, compared to the QCD prediction.}
\end{center}
\end{figure}
We observe that the $N$ dependence of the ratio is well described
by the functional form
\begin{equation}\label{eq:depn}
\frac{1}{1+ 0.044 \,(N-1)}\;,
\end{equation}
where,
since the first
moment of the non-singlet distribution should be exactly
given by the theory (because of charge conservation),
we normalize to one the extrapolation of the data to $N=1$.

We can only speculate about the possible origin of the discrepancy and the
form of the coefficient of $(N-1)$ in Eq.~(\ref{eq:depn}).
Assuming that we are dealing with a non-perturbative correction
to the coefficient function of the form
\begin{equation}\label{eq:cfpc2}
1+\frac{C(N-1)}{q^2}\;,
\end{equation}
this would lead to the extra factor
\begin{equation}
      \frac{1+\frac{C(N-1)}{M_Z^2}}{1+\frac{C(N-1)}{M_\Upsilon^2}}\;,
\end{equation}
(where  $M_Z$ and $M_\Upsilon$ are the $Z^0$ and $\Upsilon(4S)$ mass)
to be applied to our prediction for the ALEPH data.
For $C=5~{\rm GeV}^2$ we reproduce the behaviour of
Eq.~(\ref{eq:depn}).
In Ref.~\cite{Dasgupta:1996ki}, on the basis of a calculation
of infra-red renormalon effects, a $1/q^2$ power correction is found,
with an $N$ dependence marginally compatible with~(\ref{eq:cfpc2}).
No $1/E$ correction is found.
Ref.~\cite{Beneke:1997sr} also predicts a leading $1/E^2$ power correction.
 However, the
$C \approx 5~{\rm GeV}^2$ coefficient would appear to be somewhat too
large\footnote{If we believe that it is the maximum meson energy, not $E$,
that controls power effects, than we would have $C\approx 1~{\rm GeV}^2$,
a more acceptable value.}.
Alternatively, if we admitted the existence of corrections to the coefficient
functions of the form
\begin{equation}\label{eq:cfpc1}
1+\frac{C(N-1)}{E}\;.
\end{equation}
then we would find $C\approx 0.52~{\rm GeV}$, a much more acceptable value.
We observe that a form
\begin{equation}
\left(1+\frac{C}{E}\right)^{N-2} \approx 1 + \frac{C(N-2)}{E}
\end{equation}
was required
in Ref.~\cite{Nason:1994xx} to fit light-hadron fragmentation data.

Demonstrating the absence (or the existence) of $1/E$ corrections
in fragmentation functions would be a very interesting result, since
it would help to validate or disprove renormalon-based predictions.
Unfortunately, the low precision of the available data  does not allow, at the
moment, to resolve this issue.

We would like to remark that the discrepancy between the CLEO/BELLE and
ALEPH data exclusively depends upon the evolution between the $\Upsilon(4S)$
and $Z^0$ energies. The method we used to describe the CLEO/BELLE data
(i.e.\ the perturbative calculation of the fragmentation function,
the Sudakov effects in the initial conditions and the parametrization
of the non-perturbative part) does not affect the conclusions of the
present section.
 In fact, we can simply compute the ratio of the moments of the
inclusive $D^{*+}$ (ISR corrected) distribution at CLEO/BELLE and ALEPH,
and compare it to the theoretical prediction.
The result of this comparison (where we have used, for simplicity,
BELLE data only) is displayed in Fig.~\ref{fig:ALEPHoBELLE}.
\begin{figure}[htb]
\begin{center}
\epsfig{file=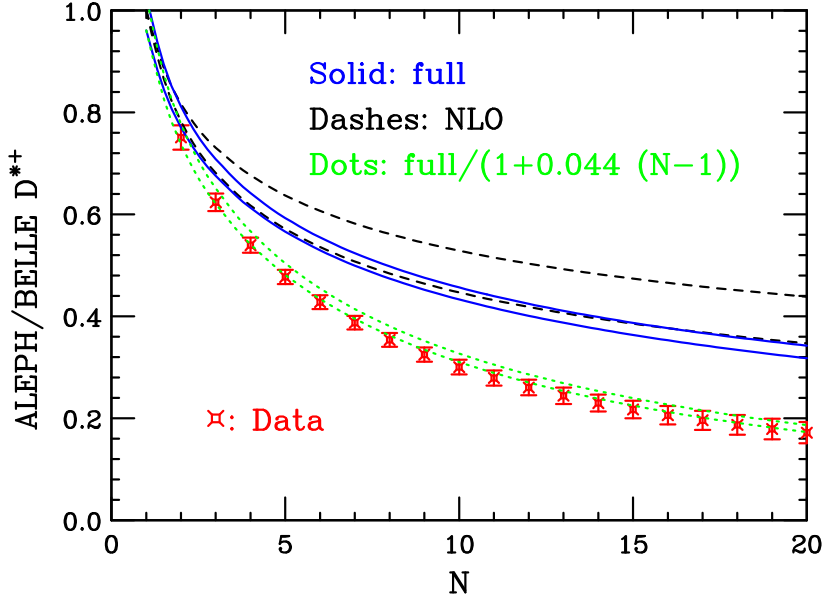,width=0.8\textwidth}
\caption{\label{fig:ALEPHoBELLE}
The ratio of ALEPH and BELLE moments for the $D^{*+}$ fragmentation
function, compared to QCD evolution.  The solid band is obtained with
QCD NLO evolution and Sudakov effects in the coefficient functions,
while the dashed bands is NLO evolution only.
The bands are obtained by setting $\mu_{Z/\Upsilon}=\xi M_{Z/\Upsilon}$
and varying $1/2<\xi<2$.}
\end{center}
\end{figure}
The curves are given by
\begin{equation}\label{eq:evfac}
\frac{\sigma_\sQ(N,M_Z^2,m^2)}{\sigma_\sQ(N,M_\Upsilon^2,m^2)} = 
\frac{\bar{a}_q(N,M_Z^2,\mu_Z^2)}{
1+\as(\mu_{Z}^2)/\pi}
\;
E(N,\mu_Z^2,\mu_{\Upsilon}^2) \;
\frac{1+\as(\mu_{\Upsilon}^2)/\pi}{
\bar{a}_q(N,M_\Upsilon^2,\mu_{\Upsilon}^2)}
\end{equation}
where $\mu_Z$ and $\mu_\Upsilon$ are the factorization scales and
the evolution factor $E$ is given in Eq.~(\ref{eq:evoqq}).
Notice that low-scale effects, both at the heavy
quark mass scale and at the non-perturbative level, cancel completely in
this ratio, making its prediction entirely perturbative.
For $\bar{a}_q$, in the NLO results (dashed lines), we have used
\begin{equation}
\bar{a}_q(N,q^2,\mu^2)=1+\asb(\mu^2)\, a_q^{(1)}(N,q^2,\mu^2)\;,
\end{equation}
while for the full result (solid lines) we have included
the NLL resummation of soft gluon emission in the coefficient functions
\beqn
\label{eq:coeffun_sud_noexp}
\bar{a}_q(N,q^2,\mu^2)&=&\Delta_q^S(N,q^2,\mu^2) \nonumber\\
&&\times \left\{  1+\asb(\mu^2) \left[a_q^{(1)}(N,q^2,\mu^2)
-\lq\Delta_q^S(N,q^2,\mu^2)\rq_{\as}\right]\right\}. \phantom{aaa}
\eeqn
The definitions of $a_q^{(1)}$ and $\Delta_q^S$ are
given in Sections~\ref{sec:Collinear_logarithms}
and~\ref{sec:Soft_logarithms}. 
We have set $\mu_{Z/\Upsilon}=\xi M_{Z/\Upsilon}$
with $\xi=0.5,2$ to plot our bands. 
As we can see from the figure, the rather large scale uncertainty
displayed by the NLO result is much reduced when Sudakov effects
are included. In both cases, however, the data clearly undershoot
the pure QCD prediction, being instead compatible with the inclusion
of the correction factor~(\ref{eq:depn}) (dotted lines).
We have also checked that our full result is essentially unchanged
if, instead of formula~(\ref{eq:coeffun_sud_noexp}), we use
the fully exponentiated formula~(\ref{eq:a_Q^res}).
Furthermore, the change of variable given in Eq.~(\ref{eq:N_tilde})
to deal with the Landau pole has very little impact on our curves.
Using the very large value $\LambdaQCD^{(5)}=0.3\,$GeV would lower the
theoretical predictions by no more than 11\%{} for $N\leq 20$,
very far from explaining the observed effect.

The deconvolution of ISR effects, that hardens the $\Upsilon(4S)$ data, but is
insignificant on the $Z^0$, widens the discrepancy.
However, if we did not apply the deconvolution, the effect would
still be partially visible.

Because of the relatively low energy of the data on the $\Upsilon(4S)$,
it is legitimate to wonder whether charm-mass effects could be responsible
for the discrepancy between LEP and $\Upsilon(4S)$ data.
We have not included mass effects in the present calculation.
However, in Ref.~\cite{Nason:1999zj}, mass effects in charm
production on the $\Upsilon(4S)$ where computed at order $\as^2$,
and found to be small. We thus believe that it is unlikely that mass effects
could play an important role in explaining this discrepancy.

\section{\boldmath{$B$} mesons data fits on the \boldmath{$Z^0$}}
\label{sec:bfit}
The same framework that yields good fits to $D$ meson production data
can also be used to describe $B$ meson production on the
$Z^0$. Accurate data have been published by the ALEPH~\cite{Heister:2001jg},
OPAL~\cite{Abbiendi:2002vt} and SLD~\cite{Abe:2002iq} Collaborations. 
Preliminary
data are available from DELPHI~\cite{Baker:2002,Ben-haim:2004kn}.
We find that, to describe $B$ production, the $\delta(1-x)$ term in
Eq.~(\ref{eq:threepar}) is in fact not needed, i.e.\ the $c$ parameter
tends to become very large in the fitting procedure.
In this limit the form of Eq.~(\ref{eq:threepar}) becomes
a two parameter form, coinciding with that of Ref.~\cite{Colangelo:1992kh}.
In Fig.~\ref{fig:Bfit} we show the result of a simultaneous fit
to ALEPH and SLD data.
\begin{figure}[htb]
\begin{center}
\epsfig{file=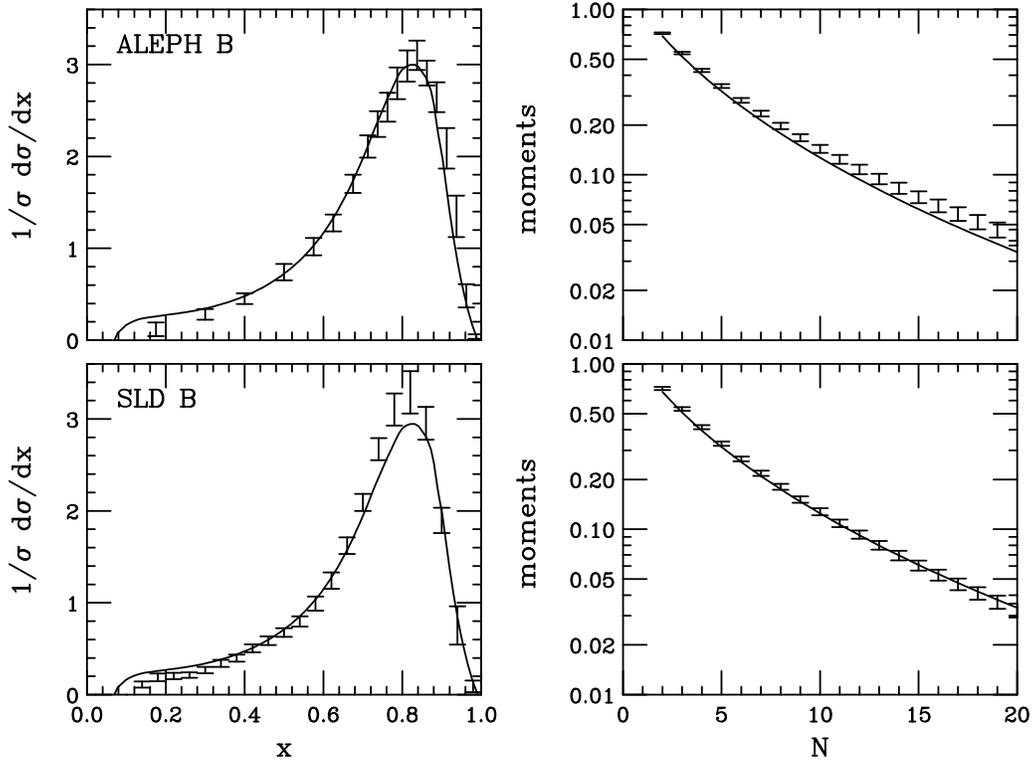,width=\textwidth}
\caption{\label{fig:Bfit}
Fit of the fragmentation function for $B$ production
together with ALEPH (upper) and SLD (lower) data.}
\end{center}
\end{figure}
In Fig.~\ref{fig:OPAL-DELPHI-b} we show the same
best-fit curve together with the OPAL and DELPHI data.
\begin{figure}[htb]
\begin{center}
\epsfig{file=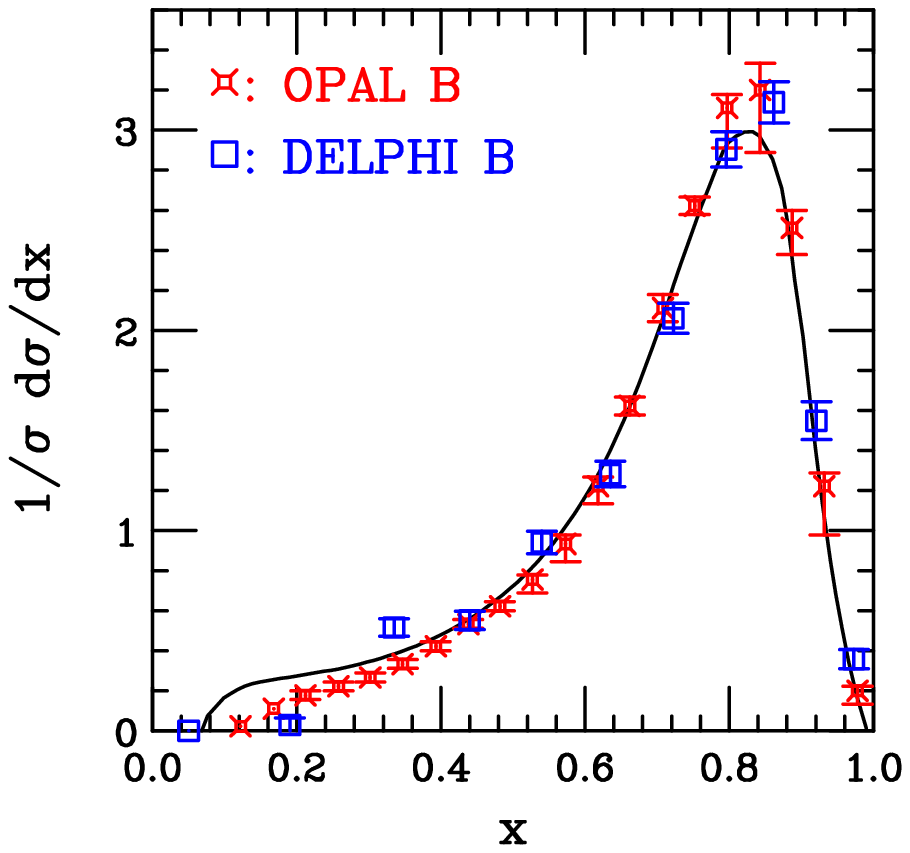,width=0.6\textwidth}
\caption{\label{fig:OPAL-DELPHI-b}
The fragmentation function fitted to ALEPH and SLD $B$ data
shown here together with OPAL and DELPHI data.}
\end{center}
\end{figure}
The fit yields $a=24\pm 2\,, b=1.5\pm 0.2$ with a $\chi^2=43$ for SLD and
$51$ for ALEPH for 21 and 19 data points respectively
 (for these data, bin-to-bin correlations provided by the
experimental Collaboration were also taken into account). 
We did not attempt to fit the OPAL data together, since it was not clear
to us how to handle the asymmetric, correlated systematic errors given
by OPAL. However, it is clear from Fig.~\ref{fig:OPAL-DELPHI-b} that also this
data set, as well as the preliminary data from DELPHI,
is well described by the fit\footnote{Note that, while the ALEPH 
set refers specifically to $B$ mesons, the SLD, DELPHI 
and OPAL data are for all weakly
decaying $b$-flavoured hadrons. 
The two quantities could therefore be slightly different,
due to the small fractions of $B_s$ and $B$ baryons
(10\%{} each, see Ref.~\cite{Eidelman:2004wy}). For an example of a 
quantitative estimate see Eq.~(5.10) of Ref.~\cite{Ben-haim:2004kn}.}.

%

\section{Moment-space fits and power corrections}{\label{sec:moment}}
The fits presented so far have been performed on the measured $x$-space
distributions, and they were aimed at providing an accurate description of
all the experimental data. This has required a flexible parametrization for
the non-perturbative fragmentation function, leading to the choice of the
three-parameter form given in Eq.~(\ref{eq:threepar}). The data are fitted
well in the large-$x$ region, so that all moments of the fragmentation
function are also well reproduced. This is important, since, as noted in
Refs.~\cite{Frixione:1998ma,Nason:1999ta}, heavy-flavour production spectra
in hadronic collisions are determined by a few Mellin moments (usually in the
range $N=2,\ldots{}6$) of the non-perturbative fragmentation function.  This
property was successively exploited in
Refs.~\cite{Cacciari:2002pa,Cacciari:2003uh,Cacciari:2003zu,Cacciari:2005rk}
for predicting bottom and charm spectra in $p\bar p$ collisions.
Inaccuracies in the description of the large-$x$ region in $e^+e^-$
annihilation could therefore lead to large errors in the moments that are
relevant to the hadroproduction of heavy quarks.  Conversely, in the
framework of heavy-flavour production, an accurate fit in $x$-space is
unnecessary, as long as the moments are well fitted in the relevant
range. For this purpose, it is therefore convenient and sufficient to use for
the non-perturbative fragmentation function one-parameter functional forms
that are commonly found in the
literature~\cite{Kartvelishvili:1978pi,Peterson:1983ak,Braaten:1994bz}.  In
the following discussion, we will focus upon these one-parameter forms, and in
particular on the one of Ref.~\cite{Kartvelishvili:1978pi}
\begin{equation}\label{eq:Kartvelishvili}
D_{\rm NP}(x) = (\alpha+1)(\alpha+2) x^\alpha (1-x)\,.
\end{equation}

\begin{figure}[htb]
\begin{center}
  \epsfig{file=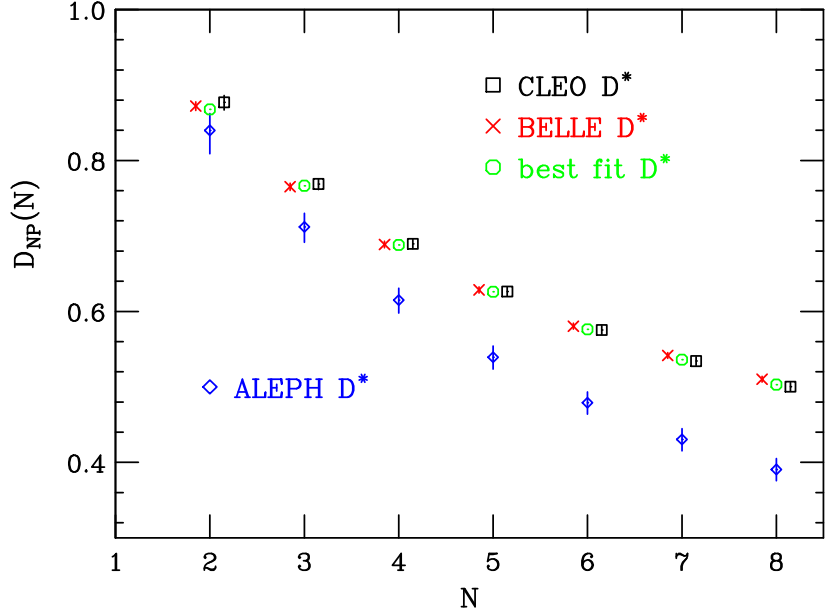,width=0.8\textwidth}
\caption{\label{fig:COMP-moments-Dstar+}
Moments of the non-perturbative component $D_{\rm NP}(N)$
extracted from $e^+e^-$ $D^*$ data, and those of the fitted 
non-perturbative fragmentation function~(\protect\ref{eq:threepar}) 
with the parameters of Table~\protect\ref{tab:fitdstar}.}
\end{center}
\end{figure}

It is important to stress that the choice of a specific parametrization
like this one is exclusively a matter of convenience,
aimed at easing the transfer of the non-perturbative information from
$e^+e^-$ collisions to other processes. One can either choose a
different functional form, or even analyze the data in terms of
non-perturbative moments $D_{\rm NP}(N)$ only. 
In Fig.~\ref{fig:COMP-moments-Dstar+}, we show the moments $D_{\rm NP}(N)$
extracted from $e^+e^-$ data. The points in the figure are obtained
by taking the experimental values of the moments of the fragmentation
function together with their errors, divided by the pure perturbative component
of the fragmentation function,
computed with our default parameters\footnote{The perturbative
fragmentation function for ALEPH is computed using the
non-singlet component only, since gluon-splitting contributions have been
subtracted from the published experimental distribution.}.
In the figure we also show the non-perturbative component given by the
form~(\ref{eq:threepar}), with the parameters taken from
Table~\ref{tab:fitdstar}. 
Also evident is the poor
consistency between values obtained from data taken on the $\Upsilon(4S)$
and on the $Z^0$. This is, of course, the same
situation already observed in Section~\ref{sec:aleph}.

\begin{figure}[htb]
\begin{center}
  \epsfig{file=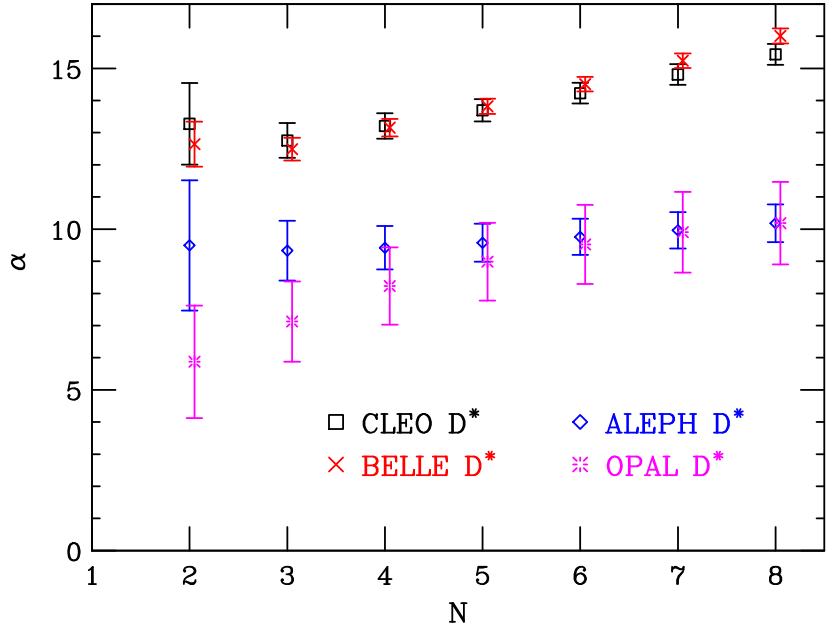,width=0.8\textwidth}
\caption{\label{fig:COMP-Dstar+}
Fits to $D^{*+}$ data for the parameter $\alpha$ of the
parametrization~(\ref{eq:Kartvelishvili}).}
\end{center}
\end{figure}

Using the Mellin transform of formula~(\ref{eq:Kartvelishvili})
\begin{equation}\label{eq:momkart}
D_{\rm NP}(N)=\frac{(\alpha+1)(\alpha+2)}{(\alpha+N)(\alpha+N+1)}\,,
\end{equation}
we can translate the moments in Fig.~\ref{fig:COMP-moments-Dstar+}
into values for $\alpha$ with the appropriately propagated error.
The results are displayed in Fig.~\ref{fig:COMP-Dstar+}.
From the figure we see that the one-parameter form~(\ref{eq:Kartvelishvili})
does not describe perfectly the whole shape, as shown by the non-constancy of
$\alpha$ extracted from different moments.
However, to a good degree of approximation a single value
of $\alpha$ can describe all the moments up to $N \simeq 6$ or so. This
is enough for the purpose of using the fitted function for convoluting
a $p_T$ distribution in hadronic collisions.

\subsection[Scaling property: from $D$ to $B$ mesons]
{Scaling property: from \boldmath{$D$} to \boldmath{$B$} mesons}
\label{sec:scaling}
Several theoretical arguments in
QCD~\cite{Nason:1997pk,Jaffe:1993ie,Randall:1994gr,Cacciari:2002xb} predict
for the heavy quark non-perturbative fragmentation function the behaviour
\begin{equation}
D_{\rm NP}(N) = 1-(N-1)\Lambda/m + {\cal O}(\Lambda^2/m^2)\,,
\end{equation}
where $\Lambda$ is a hadronic scale, and $m$ is the mass of the heavy quark.
If $D_{\rm NP}(N)$ depends
upon a single parameter, its value can be linked to the ratio
$\Lambda/m$. 
For example, in the case of the form~(\ref{eq:momkart})
the series expansion in powers of $1/\alpha$ is given by 
\begin{equation}
D_{\rm NP}(N) = 
 1 -  (N-1)\frac{2}{\alpha} +
 {\cal
O}\left(\frac{1}{\alpha^2}\right) \; .
\end{equation}
Reinterpreting $\alpha \to 2\,m/\Lambda$ we will be able to check the
behaviour of the leading power correction.
Using $\Lambda \sim 300$~MeV, $m_c \simeq 1.5$~GeV and
$m_b \simeq 4.75$~GeV, one expects to find $\alpha_D \sim 10$ and
$\alpha_B
\sim 30$ when fitting $D/D^*$ and $B$ mesons
respectively.
Whatever the exact values are,
it will always be possible to test for the predicted scaling law  
\begin{equation}
\frac{\alpha_B}{\alpha_D} \simeq \frac{m_b}{m_c} \sim 3 \; .
\end{equation}

\begin{figure}[htb]
\begin{center}
  \epsfig{file=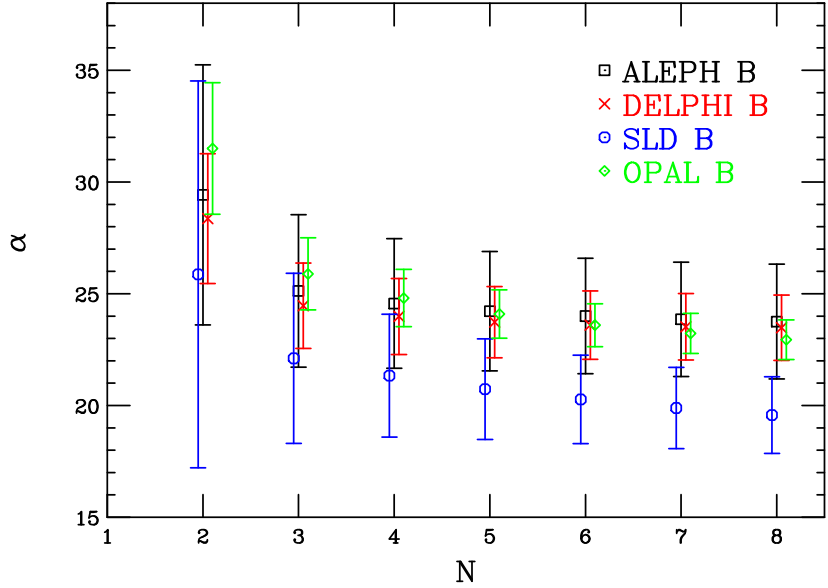,width=0.8\textwidth}
\caption{\label{fig:COMP-B}
Fits to weakly decaying $B$'s data for the parameter $\alpha$ 
of the parametrization of Ref.~\cite{Kartvelishvili:1978pi}.}
\end{center}
\end{figure}

To this end, we extract the value of $\alpha$ for $B$ meson production at
$Z^0$ energy.  In Fig.~\ref{fig:COMP-B} we show the fits to the four available
data sets.  All the data appear consistent with each other. Within fairly
large uncertainties (resulting from the non-constancy of $\alpha$ through the
fits to different moments, and the discrepancy between the determination of
$\alpha_D$ at $M_{\Upsilon}$ and $M_Z$) we can see that the expectations are
largely fulfilled, leading to an $\alpha_B/\alpha_D$ ratio of order 1.5 to 3.
The tendency for values smaller than $m_b/m_c = 3.17$ might be a consequence
of a number of factors, like the $B$ data being for ``weakly decaying''
$B$'s, and therefore generally softer than the leading $B^*$ and $B^{**}$, or
the mass entering the power corrections being closer to the meson mass rather
than the quark mass.

\begin{figure}[htb]
\begin{center}
  \epsfig{file=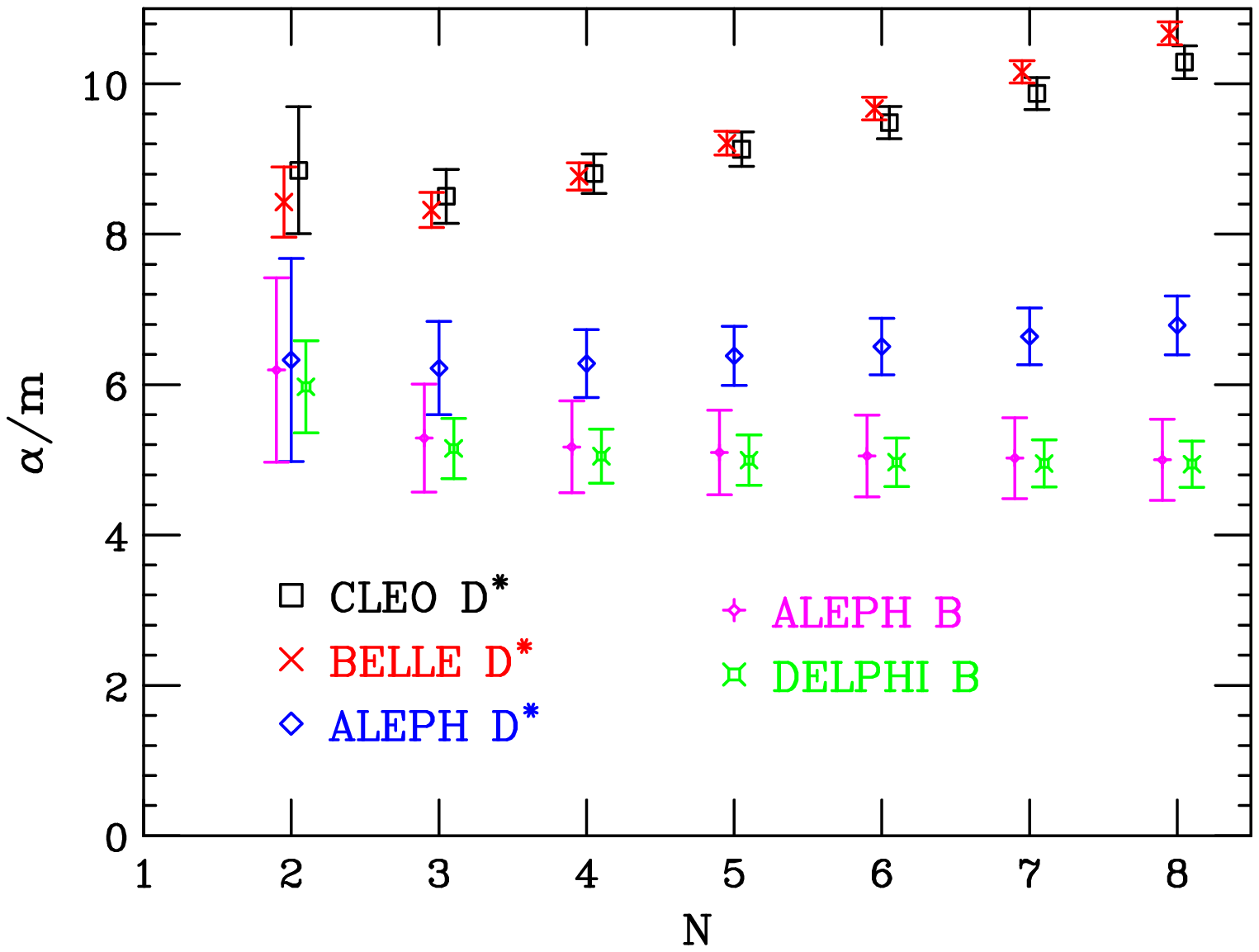,width=0.8\textwidth}
\caption{\label{fig:b-c-scal}
Values of $\alpha/m$ for $D^*$ and $B$ mesons as a function
of $N$.}
\end{center}
\end{figure}

It is also worth noting that, given a value for $\alpha_B \simeq 25$, 
the expectations for the
value of the ratio are much better fulfilled if we use the $\alpha_D$ fitted
at the $Z$ energy ($\alpha_D \simeq 9$) 
rather than the one fitted at the $\Upsilon(4S)$ energy ($\alpha_D \simeq 14$), as 
shown in Fig.~\ref{fig:b-c-scal}.
This result mildly supports the view
that large non-perturbative corrections may affect the $\Upsilon(4S)$
data.

The results of Figs.~\ref{fig:COMP-moments-Dstar+} and~\ref{fig:COMP-Dstar+}
are also summarized in Table~\ref{tab:bigtable}, together with
similar results for the one-parameter forms of Ref.~\cite{Braaten:1994bz}
and with the popular PSSZ form~\cite{Peterson:1983ak}\footnote{The values of
  $\epsilon$ that 
we find in this case are about one order of magnitude smaller than
those usually extracted from Monte Carlo simulations.
They lead, therefore, to a harder non-perturbative fragmentation
function and hence to larger rates in hadronic collisions.
It is moreover worth noting that the use of the PSSZ
fragmentation function in a context where non-perturbative corrections
are expected to scale like $1 - (N-1)\Lambda/m$  
is inconsistent. In fact,
while the $\epsilon$ parameter can be interpreted as being of order
$\Lambda^2/m^2$~\cite{Peterson:1983ak}, the series expansion of its 
Mellin transform can be shown to be
\begin{equation}
D_{\rm NP}(N) = 1 +{{2(\log\epsilon -1)(N-1) +
4N(\psi^{(0)}(N)+\gamma_E)}\over{\pi}}\sqrt{\epsilon} + {\cal
O}(\epsilon) \; .
\end{equation}
The presence of the $\log\epsilon$ term in the coefficient of the term
linear in $\sqrt{\epsilon}$ does not allow to interpret it as a simple
$\Lambda/m$ power correction.
}.
Results for $B$ mesons are also shown.

\begin{table}
\begin{sideways}
\begin{minipage}{\textheight}
\tiny
\begin{tabular}{|l|c|c|c|c|c|c|c|}
\hline
$N$ & 2 & 3 & 4 & 5 & 6 & 7 & 8 
\\
\hline
\multicolumn{8}{|c|}{$\sigma_\sQ(N,q^2,m^2) = \langle x^{N-1}\rangle_{\rm pQCD}$} \\
\hline
$c$ @ 10.58 GeV &
0.7359 &
0.5749 &
0.4601 &
0.3778 &
0.3167 &
0.2698 &
0.2331 \\
$c$ @ 91.2 GeV (NS) &
 0.5858 &
 0.3937 &
 0.2843 &
 0.2151 &
 0.1683 &
 0.1345 &
 0.1107 \\
$c$ @ 91.2 GeV (full) &
 0.5954 &
 0.3988 &
 0.2860 &
 0.2158 &
 0.1686 &
 0.1353 &
 0.1108 \\
$b$ @ 91.2 GeV &
 0.7634 &
 0.6280 &
 0.5309 &
 0.4590 &
 0.4033 &
 0.3587 &
 0.3222 \\
\hline
\multicolumn{8}{|c|}{Experimental data (norm. to one)} \\
\hline
BELLE $D^{*+} \to D^0$  (ISR corr.) &
0.6418 $\pm$  0.0042 &
0.4399 $\pm$  0.0028 &
0.3169 $\pm$  0.0020 &
0.2375 $\pm$  0.0015 &
0.1838 $\pm$   0.0012 &
0.1462 $\pm$  0.0010 &
0.1189 $\pm$  0.0009 \\
ALEPH $D^{*+}$ (ISR corr.)&
0.4920 $\pm$ 0.0152 &
0.2803 $\pm$ 0.0075 &
0.1748 $\pm$  0.0047 &
0.1160 $\pm$  0.0033 &
0.0806 $\pm$  0.0025 &
0.0582 $\pm$ 0.0020 &
0.0432 $\pm$  0.0016 \\
ALEPH $B$ &
0.7163 $\pm$  0.0085&
0.5433 $\pm$ 0.0097 &
0.4269 $\pm$  0.0098 &
0.3437 $\pm$  0.0096 &
0.2819 $\pm$  0.0094&
0.2345 $\pm$  0.0091&
0.1975 $\pm$  0.0087 \\
\hline
\multicolumn{8}{|c|}{$D_{\rm NP}(N) = \langle x^{N-1}\rangle_{\rm NP}$} \\
\hline
CLEO $D^{*+}$ &
$0.877^{+0.009}_{-0.010}$ & 
$0.769^{+0.007}_{-0.007}$ & 
$0.690^{+0.006}_{-0.006}$ & 
$0.626^{+0.006}_{-0.006}$ & 
$0.576^{+0.006}_{-0.006}$ & 
$0.534^{+0.006}_{-0.006}$ & 
$0.500^{+0.006}_{-0.006}$ \\ 
BELLE $D^{*+} \to D^0$ &
$0.872^{+0.005}_{-0.006}$ & 
$0.765^{+0.005}_{-0.005}$ & 
$0.689^{+0.004}_{-0.004}$ & 
$0.629^{+0.004}_{-0.004}$ & 
$0.580^{+0.004}_{-0.004}$ & 
$0.542^{+0.004}_{-0.004}$ & 
$0.510^{+0.004}_{-0.004}$ \\ 
ALEPH $D^{*+}$ & 
$0.840^{+0.022}_{-0.031}$ & 
$0.712^{+0.018}_{-0.021}$ & 
$0.615^{+0.016}_{-0.017}$ & 
$0.539^{+0.015}_{-0.016}$ & 
$0.479^{+0.014}_{-0.015}$ & 
$0.430^{+0.014}_{-0.015}$ & 
$0.391^{+0.014}_{-0.015}$ \\ 
\hline
Tab.~\protect\ref{tab:fitdstar} and Eq.~(\protect\ref{eq:threepar}) &
 0.868 &
 0.767 &
 0.688 &
 0.626 &
 0.576 &
 0.536 &
 0.503 \\
 \hline
ALEPH $B$ &
$0.938^{+0.009}_{-0.014}$ & 
$0.865^{+0.014}_{-0.018}$ & 
$0.804^{+0.017}_{-0.020}$ & 
$0.749^{+0.019}_{-0.023}$ & 
$0.699^{+0.022}_{-0.025}$ & 
$0.654^{+0.024}_{-0.027}$ & 
$0.613^{+0.025}_{-0.029}$ \\ 
SLD $B$ &
$0.931^{+0.016}_{-0.030}$ & 
$0.850^{+0.019}_{-0.025}$ & 
$0.781^{+0.020}_{-0.024}$ & 
$0.718^{+0.021}_{-0.024}$ & 
$0.661^{+0.021}_{-0.024}$ & 
$0.610^{+0.021}_{-0.024}$ & 
$0.563^{+0.022}_{-0.024}$ \\ 
\hline
\hline
\multicolumn{8}{|c|}{KLP $\alpha$} \\
\hline
CLEO $D^{*+}$ &
$13.28\pm  1.27$ & 
$12.76\pm  0.54$ & 
$13.21\pm  0.40$ & 
$13.70\pm  0.34$ & 
$14.23\pm  0.32$ & 
$14.81\pm  0.32$ & 
$15.43\pm  0.33$ \\ 
BELLE $D^{*+} \to D^0$ &
$12.64\pm  0.70$ & 
$12.49\pm  0.35$ & 
$13.16\pm  0.27$ & 
$13.82\pm  0.24$ & 
$14.51\pm  0.23$ & 
$15.24\pm  0.23$ & 
$16.01\pm  0.23$ \\ 
ALEPH $D^{*+}$ & 
$ 9.49\pm  2.03$ & 
$ 9.33\pm  0.93$ & 
$ 9.42\pm  0.68$ & 
$ 9.58\pm  0.59$ & 
$ 9.76\pm  0.56$ & 
$ 9.96\pm  0.57$ & 
$10.18\pm  0.59$ \\ 
\hline
ALEPH $B$ &
$29.42\pm  5.82$ & 
$25.12\pm  3.41$ & 
$24.56\pm  2.90$ & 
$24.22\pm  2.67$ & 
$24.00\pm  2.58$ & 
$23.86\pm  2.56$ & 
$23.75\pm  2.57$ \\ 
ALEPH $B$, $m_b = 4.5$~GeV &
$34.32\pm  7.77$ & 
$28.26\pm  4.23$ & 
$27.34\pm  3.51$ & 
$26.73\pm  3.18$ & 
$26.34\pm  3.04$ & 
$26.07\pm  2.98$ & 
$25.87\pm  2.97$ \\ 
ALEPH $B$, $m_b = 5.0$~GeV &
$25.90\pm  4.59$ & 
$22.72\pm  2.84$ & 
$22.41\pm  2.46$ & 
$22.23\pm  2.30$ & 
$22.13\pm  2.25$ & 
$22.07\pm  2.24$ & 
$22.03\pm  2.26$ \\ 
SLD $B$ &
$25.87\pm  8.66$ & 
$22.11\pm  3.80$ & 
$21.33\pm  2.75$ & 
$20.73\pm  2.25$ & 
$20.27\pm  1.98$ & 
$19.89\pm  1.82$ & 
$19.57\pm  1.72$ \\ 
\hline
\multicolumn{8}{|c|}{BCFY $r$} \\
\hline
CLEO $D^{*+}$ &
$0.0531\pm 0.0077$ & 
$0.0610\pm 0.0036$ & 
$0.0615\pm 0.0026$ & 
$0.0611\pm 0.0021$ & 
$0.0601\pm 0.0019$ & 
$0.0587\pm 0.0017$ & 
$0.0569\pm 0.0016$ \\ 
BELLE $D^{*+} \to D^0$ &
$0.0570\pm 0.0046$ & 
$0.0628\pm 0.0025$ & 
$0.0618\pm 0.0018$ & 
$0.0604\pm 0.0014$ & 
$0.0585\pm 0.0013$ & 
$0.0564\pm 0.0011$ & 
$0.0541\pm 0.0011$ \\ 
ALEPH $D^{*+}$ &
$0.0849\pm 0.0247$ & 
$0.0936\pm 0.0125$ & 
$0.0972\pm 0.0092$ & 
$0.0988\pm 0.0080$ & 
$0.0993\pm 0.0074$ & 
$0.0990\pm 0.0072$ & 
$0.0981\pm 0.0072$ \\ 
ALEPH $D^{*+}$, $m_c = 1.3$~GeV &
$0.0470\pm 0.0238$ & 
$0.0557\pm 0.0102$ & 
$0.0594\pm 0.0074$ & 
$0.0613\pm 0.0063$ & 
$0.0621\pm 0.0059$ & 
$0.0623\pm 0.0057$ & 
$0.0619\pm 0.0057$ \\ 
ALEPH $D^{*+}$, $m_c = 1.7$~GeV &
$0.1198\pm 0.0289$ & 
$0.1288\pm 0.0146$ & 
$0.1323\pm 0.0108$ & 
$0.1336\pm 0.0093$ & 
$0.1337\pm 0.0086$ & 
$0.1329\pm 0.0084$ & 
$0.1315\pm 0.0084$ \\ 
\hline
\multicolumn{8}{|c|}{PSSZ $\epsilon$ ($\times 10^{2}$) } \\
\hline
BELLE $D^{*+} \to D^0$ &
$0.234\pm 0.032$ & 
$0.271\pm 0.019$ & 
$0.260\pm 0.013$ & 
$0.246\pm 0.010$ & 
$0.230\pm 0.009$ & 
$0.213\pm 0.007$ & 
$0.197\pm 0.007$ \\ 
ALEPH $D^{*+}$ &
$0.473\pm 0.245$ & 
$0.548\pm 0.129$ & 
$0.574\pm 0.096$ & 
$0.580\pm 0.081$ & 
$0.575\pm 0.074$ & 
$0.563\pm 0.071$ & 
$0.547\pm 0.069$ \\ 
ALEPH B & 
$0.028\pm 0.014$ & 
$0.047\pm 0.016$ & 
$0.056\pm 0.016$ & 
$0.062\pm 0.017$ & 
$0.068\pm 0.018$ & 
$0.073\pm 0.019$ & 
$0.077\pm 0.020$ \\ 
\hline
\end{tabular}
\caption{\label{tab:bigtable} 
Summary of results for the first eight moments.
The first group of lines (labelled $\sigma_\sQ(N,q^2,m^2) = \langle x^{N-1}\rangle_{\rm pQCD}$)
gives perturbative moments for $c$ and $b$ production at the $\Upsilon(4S)$ and $Z^0$ energies.
The second group (labelled ``Experimental data'') gives the measured moments when explicitly given
by the experimental Collaborations. In this case, the ISR correction (when applied)
has been taken from the right panel of Fig.~\ref{fig:isrcorr}.
The third group (labelled $D_{\rm NP}(N) = \langle x^{N-1}\rangle_{\rm NP}$) gives the
moments of the non-perturbative fragmentation function that we extracted from the data.
The last three groups report the value of the parameters of the KLP, BCFY and PSSZ parametrization
extracted from several data sets.}
\end{minipage}
\end{sideways}
\end{table}

\subsection{Implications for heavy-flavour hadroproduction}
\label{subsec:hadroproduction}
It is legitimate to ask what is the impact of the new, high precision
$\Upsilon(4S)$ data on the calculation of $D$ meson production spectra
in hadronic collisions, especially in view of the discrepancy
between $\Upsilon(4S)$ and $Z^0$ data. The question has not,
however, a straightforward answer. If the discrepancy is related to
a power suppressed effect in the $e^+e^-$ coefficient functions,
one should then privilege the $Z^0$ data, where power effects are
much reduced.
It is worth noting, however, that if we instead
use the $\Upsilon(4S)$ data, the impact on the hadronic
cross sections is quite limited.
This is clearly visible in Fig.~\ref{fig:COMP-moments-Dstar+},
where it appears
that for $N$ around 5 the $\Upsilon(4S)$ moments are higher than
the $Z^0$ ones by roughly 20\%. 
This value is directly proportional to
the $D^*$ production cross section in hadron collisions at large $p_T$.
Therefore, in this `worst case' scenario, 
having used the ALEPH data (the only accurate
ones available at the time), might have lead
Ref.~\cite{Cacciari:2003zu} to underestimate the
$D^*$ hadronic cross section
by 20\%, an uncertainty which is anyway not larger than those of purely
perturbative origin (variation of renormalization and factorization
scales) or stemming from the parton distribution functions.
\begin{figure}[htb]
\begin{center}
  \epsfig{file=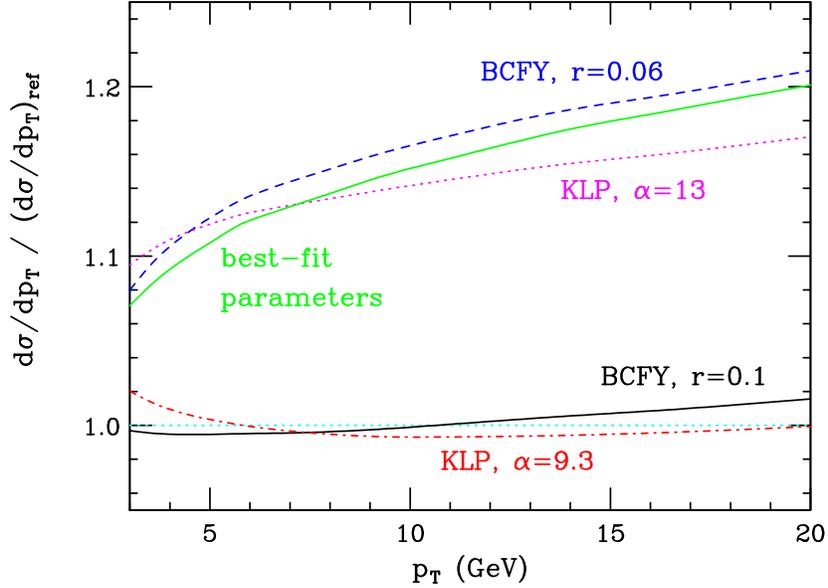,width=0.8\textwidth}
\caption{\label{fig:COMP-Tevatron}
Ratios between new evaluations of the $d\sigma/dp_T$ production cross
section of $D^*$ mesons in $p\bar p$ collisions at the Tevatron Run II
and the value originally published in~\protect\cite{Cacciari:2003zu},
$(d\sigma/dp_T)_{\rm ref}$.}
\end{center}
\end{figure}

These considerations are put on a more quantitative footing in
Fig.~\ref{fig:COMP-Tevatron}, where we plot the ratios between new
determinations of the $p_T$ distribution of $D^*$ production at the
Tevatron Run II and the central value obtained in
Ref.~\cite{Cacciari:2003zu}. The solid line, labeled `BCFY, $r$ = 0.1' 
is obtained by employing the same
non-perturbative fragmentation function and the same parameter
$r = 0.1$ as in~\cite{Cacciari:2003zu}. Its small difference
from one is essentially of perturbative origin. It is due to the
different treatment of the perturbative fragmentation function in the 
FONLL code for heavy quark hadronic production~\cite{Cacciari:1998it,Cacciari:2001td}
which, for consistency with the extraction of the non-perturbative
parameters,  has been modified to include also the  Sudakov resummation for the
initial condition 
and the large-$N$  regularization procedure described in
Eq.~(\ref{eq:N_tilde_ini}). The five other curves
are instead obtained 
with different non-perturbative forms and/or parameters relative
to the $\Upsilon(4S)$ or to the $Z^0$ results from Table~\ref{tab:bigtable}.
As expected,
using a different functional form (KLP) but a
parameter also extracted from the ALEPH data ($\alpha = 9.3$) 
returns a result very
similar to that of Ref.~\cite{Cacciari:2003zu}. On the other hand,
using determinations from CLEO/BELLE data ($\alpha = 13$) returns a larger
cross 
section, the increase being of the order predicted above and
not larger than the uncertainties of perturbative origin.

\begin{figure}[htb]
\begin{center}
  \epsfig{file=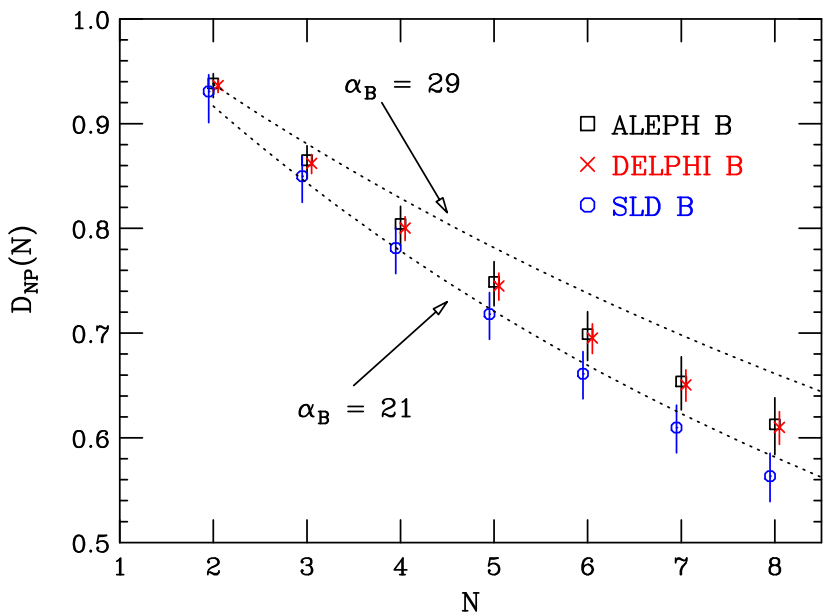,width=0.8\textwidth}
\caption{\label{fig:COMP-moments-B}
Non-perturbative moments from weakly decaying $B$'s data.}
\end{center}
\end{figure}

As far as $B$ mesons are concerned, the values for $\alpha_B$ are translated 
into non-perturbative moments
in Fig.~\ref{fig:COMP-moments-B}. The dotted band shows the values given
by two extreme choices of $\alpha_B$. One can see that, using everywhere
the value determined at $N=2$, i.e.\ $\alpha_B \simeq 29$, as done
in~\cite{Cacciari:2002pa}\footnote{In this reference, a pure NLL collinear 
resummation was used, without Sudakov resummation and large-$N$
correction factor. This does not affect, of course, the small-N region,
hence the determination of a very similar value for $\alpha_B$.}, only
overestimates the moment at $N=4$ by a few percent. Up to $N=8$ the
difference is never larger than 10\%. 
Such an uncertainty is fully acceptable when calculating the
hadronic production of $B$ mesons, given the similar or larger size of
the perturbative uncertainties and of those due to the parton distribution
functions.


\section{Conclusions}
In the present paper, we have obtained two main results.  First, we have
shown that it is possible to perform excellent fits of $D$ and $B$ meson
fragmentation spectra in perturbative QCD, using all known results on the
perturbative heavy-quark fragmentation function, and compounding them with a
simple parametrization of non-perturbative effects.  For reasons of space we
did not perform fits to available data on $D_s$ and $\Lambda_c$ production.
We can provide the corresponding results upon request.

A second striking result is the evidence of large non-perturbative effects,
visible in the relation between the $D^*$ fragmentation function at the
$\Upsilon(4S)$ and $Z^0$ energies.  It would be interesting to understand the
power law of these contributions. Their magnitude would suggest a $1/E$
scaling law. Theoretical arguments based upon infrared renormalons would
favour, instead, a $1/E^2$ behaviour. Because of the lack of precise $D$
production data in the intermediate region, it is difficult, at this point,
to discriminate between the two possibilities.  We point out, however, that,
if these non-perturbative corrections involve the coefficient functions, they
may be present also in light-hadron production, where data at intermediate
energy are available.  It is thus possible that fits to the light-hadron
fragmentation functions from $\Upsilon(4S)$ up to $Z^0$ energies may clarify
this issue.

The parametrization of the non-perturbative component of the heavy-quark
fragmentation function is also relevant for the calculation of heavy-quark
hadroproduction cross sections.  In the present work we provide various
related results, that can be used for such calculations.
\label{sec:Conc}

\section{Acknowledgments}
We wish to thank Einan Gardi for useful discussions.
We also wish to thank Giancarlo Moneti for many
discussions on CLEO data analysis. Finally we thank
Bostian Golob, Rolf Seuster and Bruce Yabsley
for providing us with BELLE data.



\providecommand{\href}[2]{#2}\begingroup\raggedright\endgroup

\end{document}